\documentclass{emulateapj}

\usepackage{graphicx}

\begin{document}

\title{The Spine of the Cosmic Web}

\author{Miguel A. ~Arag\'on-Calvo\altaffilmark{1}, Erwin Platen\altaffilmark{2}, Rien van de Weygaert\altaffilmark{2}, Alexander S. Szalay\altaffilmark{1} }
\altaffiltext{1}{The Johns Hopkins University, 3701 San Martin Drive, Baltimore, MD 21218, USA}
\altaffiltext{2}{Kapteyn Institute, University of Groningen, PO Box 800, 9700 AV Groningen, the Netherlands}

\begin{abstract}
We present the SpineWeb framework for the topological analysis of the 
Cosmic Web and the identification of its walls, filaments and cluster nodes. Based on the {\it watershed segmentation} of 
the cosmic density field, the \textit{SpineWeb }method invokes the local adjacency properties of the boundaries between the 
watershed basins to trace the critical points in the density field and the separatrices defined by them. The separatrices are 
classified into \textit{walls} and the {\it spine}, the network of filaments and nodes in the matter distribution. Testing 
the method with a heuristic Voronoi model yields outstanding results. Following the discussion of the test results, we apply 
the {\it SpineWeb} method to a set of cosmological N-body simulations. The latter illustrates the potential 
for studying the structure and dynamics of the Cosmic Web. 
\end{abstract}


\section{Introduction}
The large scale distribution of matter revealed by galaxy surveys features a complex 
network of interconnected filamentary galaxy associations. This network, which has become known as 
the {\it Cosmic Web} \citep{Bond96}, contains structures from a few megaparsecs up to tens 
and even hundreds of Megaparsecs of size. The weblike spatial arrangement of galaxies and mass into 
elongated filaments, sheetlike walls and dense compact clusters, the existence of large near-empty void 
regions and the hierarchical nature of this mass distribution -- marked by substructure over a wide range 
of scales and densities -- are its three major characteristics. Its appearance has been most dramatically 
illustrated by the recently produced maps of the nearby cosmos, the 2dFGRS, the SDSS and the 2MASS redshift 
surveys \citep[e.g.][]{Colless03,Huchra05}. 

In this paper we introduce the SpineWeb formalism for analyzing the structure and topology of the 
Cosmic Web. It identifies the sheets and filaments in the Cosmic Web, along with the large underdense 
void regions, and their mutual connection into the {\it Spine} of the cosmic matter distribution. 
The method is based on the {\it Watershed Transform} \citep[WST,][]{Beucher82}, and is largely free of user-specific 
parameters and artificial smoothing scale(s). Its output will enable the study of the physical properties and 
dynamics of the individual morphological components, along with their topology and hierarchical 
characteristics.

\subsection{the Cosmic Web}
The Cosmic Web is the most salient manifestation of the anisotropic nature of gravitational collapse, the motor 
behind the formation of structure in the cosmos \citep{Peebles80}. N-body computer simulations have 
profusely illustrated how a primordial field of tiny Gaussian density perturbations transforms into  
a pronounced and intricate filigree of filamentary features, dented by dense compact clumps at the nodes of 
the network \citep[][]{Colberg05,Springel05}. The filaments connect into the cluster nodes and act as the 
transport channels along which matter flows into the clusters. 

Fundamental understanding of anisotropic collapse on cosmological scales came with the seminal study by \citet{Zeldovich70}, 
who recognized the key role of the large scale tidal force field in shaping the Cosmic Web \citep[also see][]{Icke73}. 
The collapse of a primordial cloud (dark) matter passes through successive stages, first assuming a flattened \textit{sheetlike} 
configuration as it collapses along its shortest axis. 
This is followed by a rapid evolution towards an elongated \textit{filament} as the intermediate axis collapses and, if 
collapse continues along the longest axis, may ultimately produce a dense, compact and virialized \textit{cluster} or 
\textit{halo}. The hierarchical setting of these processes, occurring simultaneously over a wide range of scales and 
modulated by the expansion of the Universe, complicates the picture considerably. Recent state-of-the-art computer experiments 
like the Millennium simulation \citep[][]{Springel05} clearly show the hierarchical nature in which not only the clusters 
build up but also the filamentary network itself \citep[see][]{Aragon07b}. 

The Cosmic Web theory of \cite{Bond96} succeeded in synthesizing all relevant aspects into a coherent dynamical 
and evolutionary framework. It is based on the realization that the outline of the cosmic web may already be recognized in 
the primordial density field. The statistics of the primordial tidal field explains why the large scale universe looks 
predominantly filamentary and why in overdense regions sheetlike membranes are only marginal features \citep{pogosyan98}. Of key importance 
is the observation that the rare high peaks, which will eventually emerge as clusters, are the dominant agents for generating 
the large scale tidal force field: it is the clusters which weave the cosmic tapestry of filaments 
\citep[][]{Bond96,WeyEdb96,WeyBond08a}. They cement the structural relations between the 
components of the Cosmic Web and themselves form the junctions at which filaments tie up. This relates the 
strength and prominence of the filamentary bridges to the proximity, mass, shape and mutual orientation of the generating 
cluster peaks: the strongest bridges are those between the richest clusters that stand closely together and point into each 
other's direction.

The emerging picture is one of a primordially and hierarchically defined network whose weblike topology is imprinted over a wide 
spectrum of scales. Weblike patterns on ever larger scales get to dominate the density field as cosmic evolution proceeds, 
and as small scale structures merge into larger ones. Within the gradually emptying void regions, however, the topological 
outline of the early weblike patterns remains largely visible.

\subsection{Closing in on the Cosmic Web}
Despite a large variety of attempts, as yet no generally accepted descriptive framework has 
emerged for the objective and quantitative analysis of the geometry and topology of the Cosmic Web. 
The great complexity of both the individual structures as well as their connectivity, the lack of structural symmetries, its 
intrinsic multiscale nature and the wide range of densities that one finds in the cosmic matter 
distribution has prevented the use of simple and straightforward techniques.

Historically, the quantitative analysis of the Cosmic Web has been dominated by a description 
in terms of statistical measures of clustering of galaxies and matter. 
While correlation functions have been the mainstay of the cosmological analysis of large scale structure, a 
direct interpretation in terms of the patterns and texture of the Cosmic web has largely remained 
elusive. Over the years a variety of heuristic measures have been forwarded to analyze specific aspects of the spatial 
patterns in the large scale Universe, 
but only in recent years there have been attempts towards developing complete descriptors of the intricate 
spatial patterns that define the Cosmic Web. Nearly without exception these methods borrow extensively from other 
branches of science such as image processing, mathematical morphology, computational geometry and medical imaging. 

\begin{figure*}[!ht]
  \centering
  \includegraphics[width=0.49\textwidth,angle=0.0]{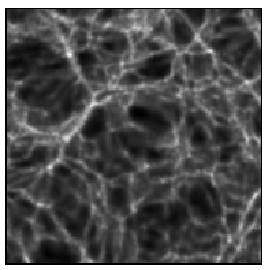}
  \includegraphics[width=0.49\textwidth,angle=0.0]{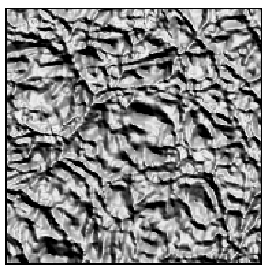}
  \includegraphics[width=0.49\textwidth,angle=0.0]{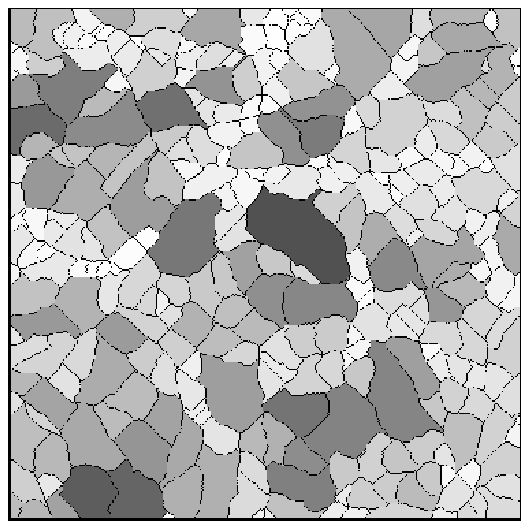}
  \includegraphics[width=0.49\textwidth,angle=0.0]{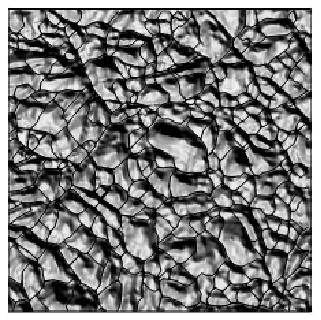}
  \caption{Top left: a slice of 100 $h^{-1}$ Mpc of side showing the density field computed from a N-body simulation. 
	   Top right: the same slice as the left panel but here showing the density field as a shaded landscape where
	      high density regions correspond to ridges while underdense regions correspond to valleys.
	      The light source used to shade the surface is located at the north-east.
           Bottom left: the 2D watershed transform computed from the 2D density field using a discrete-intensity  bucket algorithm (black contours). Each individual valley 
              is randomly colored for visualization purposes.
	   Bottom right: the same landscape as the top right panel but here we also show the watershed transform superimposed 
	      as thick black lines. The watershed contour has been slightly smoothed for visualization purposes.}
  \label{fig:density_landscape}
\vskip 0.35truecm
\end{figure*} 

Noteworthy examples include filament detection with the help of the Candy model \citep{Stoica05} and 
wavelet analysis of the Cosmic Web \citep{Martinez05}. Several methods seek to relate morphological 
features to singularities in the density field, usually invoking information on the gradient and Hessian of 
the density field, or of the tidal field \citep[see e.g.][]{Sousbie08,Aragon07a,Aragon07b,Hahn07a,Hahn07b,Bond09}. 
A classification scheme on the basis of the manifolds in the tidal field -- involving all morphological 
features in the cosmic matter distribution -- has been presented by \cite{Hahn07a,Hahn07b,Forero08}.
However, its success may depend strongly on the correct choice of the smoothing scale.
Another concept addressing the gradient and Hessian of the density field is that of the {\it skeleton analysis}, a direct 
application of Morse theory to cosmological density fields \citep[see][]{Colombi2000,Pogosyan09}. 
The skeleton formalism has been developed for the morphological analysis of the Megaparsec Cosmic Web, 
in redshift surveys like SDSS as well as in N-body simulations \citep{Novikov06,Sousbie08,Sousbie08b,Sousbie09}. 
Its present implementation refers to features identified at one single specific scale and suffers from 
Gaussian smoothing. 
The multiscale nature of the cosmic matter distribution is explicitly addressed by the Multiscale Morphology Filter, 
which is based on a scale-space analysis of the Hessian of the density field \cite{Aragon07a,Aragon07b} to 
identify cluster, filaments and sheets on the scale where they are locally most prominent. 

\subsection{Watershed and Cosmic Spine}
One technique that implicitly addresses the topology of the Cosmic Web is the Watershed Void Finder (WVF)
developed by \cite{Platen07}. The WVF is an application of the {\it Watershed Transform} for the
identification of underdense basins in the megaparsec-scale matter distribution. 
The Watershed transforms segments the density field into isolated basins and delineates the 
boundaries of cosmological voids.

The method presented in this paper is the natural extension of the Watershed Void Finder. It includes
the watershed transform into a wider context as a framework for studying both the
morphology and topology of the cosmic web and its various constituents. The result is 
the \textit{SpineWeb} method, a complete framework for the identification of voids, walls and filaments. 
Via the practical role of the watershed transform in computing the Morse complex 
it is intimately related to Morse theory, in which it finds its mathematical foundation. 
It is important to note that although Morse theory can be used to describe the same topology traced
by the SpineWeb method there is not a univocal relation between the two as the SpineWeb is based on
the observed properties of the Cosmic Web.

An important aspect of our method is that it is an intrinsically scale-free 
method, starting from a scale-free reconstruction of the density field. We use the 
DTFE method of \cite{Schaap00}, which guarantees an optimal and unbiased representation 
of the hierarchical nature and anisotropic morphology of cosmic structure \citep[see][for 
an extensive description]{WeySchaap09}. Having guaranteed the capability of invoking a 
full scale-free {\it Scale-space} representation of cosmic structure, our watershed procedure not 
only traces the outline of filaments and sheets, but may also be extended towards doing so 
over a range of scales in order to address their hierarchical structure.  

\subsection{Outline}
The principal rationale behind the SpineWeb analysis of the cosmic matter 
distribution is the interest in relating the geometry of the matter and galaxy distribution in a more 
meaningful fashion to the underlying dynamical evolution. One particular aspect of this 
dynamically motivated disentanglement of structure is the attempt to identify various 
evolutionary stages of the tidally induced anisotropic collapse of structure in 
the Universe. 

In this paper we will focus specifically on the description of the basic 
SpineWeb formalism, confined to a density field sampled on a regular grid. 
We start by discussing the topological background of our study in 
section~\ref{sec:watershed}, focusing on the watershed transform, its connection 
to the general context of Morse theory and the related issues of practical 
interest to the SpineWeb formalism. The overall cosmological background of 
the structure, formation and dynamics of the Cosmic Web follows in 
section~\ref{sec:cosmicweb}, amongst others to establish the link between 
the structures identified by the SpineWeb technique and the filamentary 
identity of tidal bridges in the theory of the Cosmic Web \citep{Zeldovich70,Bond96}. 
The technical aspects of the SpineWeb formalism are outlined in some detail in 
section~\ref{sec:spineweb}. Subsequently, the formalism is tested by applying it 
to two different classes of spatial particle distributions. The first testbed 
concerns two simple heuristic Voronoi clustering models which model aspects of 
cellular and/or weblike spatial distributions. Visual and quantitative 
tests are described in section~\ref{sec:voronoi}. The operation of SpineWeb in 
a more realistic setting of a $\Lambda$CDM simulation is the subject of 
section~\ref{sec:lcdm}. In this section we also stress the fundamental 
differences between a structural selection based on density thresholds or 
one based on topological criteria. To illustrate the potential for analyzing 
cosmological structures, in section~\ref{sec:analysis} we shortly describe three 
quantitative measures for the matter distribution in the $\Lambda$CDM 
simulation we used for testing. Finally, section~\ref{sec:conclusions} 
summarizes our results, and discusses prospects and further developments 
of the SpineWeb formalism.


\bigskip
\section{Watershed Segmentation of the Cosmic Web}
\label{sec:watershed}
When studying the topological and morphological structure of the cosmic matter distribution 
in the Cosmic Web, it is convenient to draw the analogy with a landscape (see 
fig.~\ref{fig:density_landscape}, top row). \textit{Valleys} represent 
the large underdense voids that define the cells of the Cosmic Web. Their 
boundaries are \textit{sheets} and \textit{ridges}, defining the network 
of walls, filaments and clusters that defines the Cosmic Web (cf. top panels 
fig.~\ref{fig:density_landscape}). 

\medskip
\subsection{the Watershed Transform}
The watershed transform (WST) is one of the most common methods used in Image analysis 
for segmenting images into distinct patches and features. It is a concept defined within the 
context of mathematical morphology, and was first introduced by \cite{BeuLan79}. 
The basic idea behind the WST stems from geophysics, where it is used to 
delineate the boundaries of separate domains, i.e. {\it basins} into which yields of e.g. rainfall will collect. 
The watershed transform is formed by the ridges and sheets surrounding the watershed basins and includes a subset 
of all the critical points in the density field. 

The word {\it watershed} finds its origin in the analogy of the procedure with that of 
a landscape being flooded by a rising level of water. Suppose 
we have a surface in the shape of a landscape (cf. top right panel, fig.~\ref{fig:density_landscape}). 
The surface is pierced at the location of each of the minima. As the water-level rises a growing fraction 
of the landscape will be flooded by the water in the expanding basins. Ultimately basins will meet 
at the ridges defined by {\it saddle-points} and {\it maxima} in the density field. The final 
result of the completely immersed landscape is a division of the landscape into individual 
cells, separated by {\it ridge dams} (see left bottom panel fig.~\ref{fig:density_landscape}). 

\subsection{A watershed search for voids}
The watershed transform was first introduced in a cosmological context as an objective technique 
to identify and outline voids in the cosmic matter and galaxy distribution \citep{Platen07,Platen09}. 
Following the density field-landscape analogy, the Watershed Void Finder (WVF) method identifies 
the underdense void patches in the cosmic matter distribution with the watershed basins. The method is 
parameter free in case there is no noise in the data. 

A major advantage of the WVF method is its independence of assumptions on the shape and size of voids 
(see \citep{Colberg08} for a comparison of its performance with a variety of void finding algorithms). 
Sharing this virtue with a similar tessellation-based void finding method, ZOBOV \citep{Neyrinck08}, WVF is 
particularly suited for the analysis of the hierarchical void distribution expected in the commonly 
accepted cosmological scenarios. 

\begin{figure*}[!ht]
  \includegraphics[width=0.49\textwidth,angle=0.0]{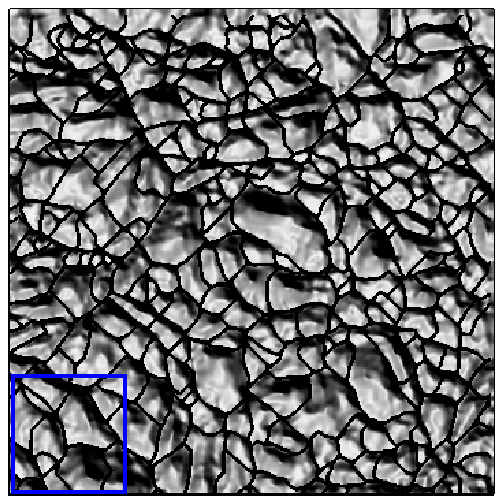}  
  \includegraphics[width=0.49\textwidth,angle=0.0]{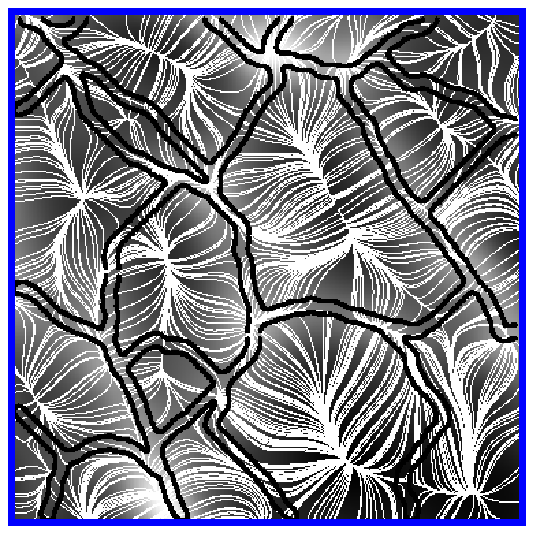}  	
	\caption{Left: A slice of the density field shown as a shaded landscape with the watershed
		lines superimposed as black lines. Right: The zoomed area in the blue square of the
		left panel showing the slope lines (white lines) superimposed on the density field
		(gray background). The contour of the watershed transform is delineated by the thick
		black lines.}
  \label{fig:landscape_flow}
  \vskip 0.25truecm
\end{figure*} 

\subsection{Watersheds and Landscape Gradients}
Extrapolating its application to other areas of interest, the implementation of the watershed 
transform may also be seen as a practical instrument for the segmentation of surfaces and volumes  
on the basis of the topological structure of the ``landscape'' $f({\bf x})$. To trace the 
topological structure of a field $f({\bf x})$, we need to investigate the structure of the gradient 
field of the landscape, $\nabla f$ (for an excellent introduction to computational topology, we 
refer to \citet{Edelsbrunner10}).

\subsubsection{Gradient Field and Integral Lines}
The gradient delineates a smooth vector field, which vanishes at critical points, 
\begin{equation}
\nabla f({\bf x_k})\,=0\,,
\end{equation}

The {\it integral lines} or {\it slope lines} represent the flow along the gradient field $\nabla f$ 
between the critical points. On the basis of these connections one may infer a variety of spatial 
segmentations \citep[see e.g.][]{Cayley1859,Maxwell1870,Eberly94,Furst02,Edelsbrunner03a,Edelsbrunner03b,
Danovaro03,Gyulassy05}. One particular segmentation is the watershed transform, which segments the 
landscape $f$ into regions of uniform local gradient behavior: the {\it watershed basin} $j$ consists of the 
collection of points ${\bf x}$ that are closer in topographic distance $\mathcal{T}({\bf x},{\bf y}_j)$, 
\begin{equation}
\mathcal{T}({\bf x},{\bf y}_j)\,\equiv\,{\rm inf} \int_{\Gamma} |\nabla f(\gamma(s))| ds\,.
\label{eq:topdist}
\end{equation} 
to the defining minimum ${\bf y}_j$ of the basin than to any of the other minima. In this definition the integral 
is the {\it pathlength} along the {\it integral line}, the line along whose path the tangent at each point is 
parallel to the local gradient $\nabla f$. The {\it watershed} itself then consists of the ridge lines that 
delimit the boundaries between basins in the terrain. 

An illustration of the close link between the gradient field and structural features in the Universe is 
offered by the righthand panel of fig.~\ref{fig:landscape_flow}. The image shows that the integral lines 
that define the boundaries of adjacent valleys are in fact the \textit{watersheds}. It also reveals the 
intimate relationship between the critical points in the flow field and the nodes, filaments and voids in 
the landscape: maxima are found at nodes of the weblike network of watershed ridges, minima at the centers 
of the void cells, while saddle points are to be found at key locations along the ridges. Following this view, 
we see that the watershed lines are the set of slope lines emanating from saddle points and connecting to a 
local maximum or minimum. Within this framework, saddle points have the crucial function of defining the sheets 
and filaments in the density field through their connection to the maxima via the integral lines. 

Note that because the image in fig.~\ref{fig:landscape_flow} is a slice through a three-dimensional field, the 
identification between the structural elements and the critical points in the image is not entirely unequivocal 
(see below). Nonetheless, the principal observation is that the resulting weblike segmentation of space, and 
the corresponding boundary manifolds, contain the full information on its topological structure marked by sheets, 
filaments and nodes. 

\subsection{Morse theory}
\label{sec:morse}
The vast majority of applications of the watershed transform concern the interior of the segmented regions. However, 
it is straightforward to extend its focus to other morphological components of the Cosmic Web, towards the delineation of 
the network of overdense ridges and walls which form the boundary manifolds of the cosmic density landscape. 

This can be directly appreciated by noting the close relation between the definition of the watershed transform and the 
more formal concept of the Morse complex. Morse theory is the mathematical 
framework for the analysis of the topological structure of manifolds, by relating it to smooth, C$^2$-differentiable, 
functions defined on those spaces. Central to Morse theory are the location and nature of the critical points -- minima, 
maxima and saddle points -- and their mutual connection via the gradient-based {\it integral lines}. These determine the 
morphological features of the functional surface. 

Even though there are some differences between the two \citep[see e.g.][]{Gyulassy08}, the close similarity between 
the definition of the watershed transform and the concepts of Morse theory indicates that 
the computation of the watershed transform may be used as an efficient means of computing the various structural 
elements in a landscape dissected along the lines of Morse theory \citep{Morse1934,Milnor1965}. 

In a cosmological context, the {\it skeleton} formalism \citep{Novikov06,Sousbie08,Sousbie08b,Sousbie09} is 
also based on Morse theoretical concepts, via the gradient and/or Hessian of the
density field. The approach followed in our SpineWeb procedure involves the specific application of the
watershed transform for the analysis and description of the
topology of the Megaparsec Universe, following our introduction of the concept
in the context of cosmic density field analysis \citep{Platen07}. Implicitly, this results in a pseudo-Morse segmentation
\citep[see e.g.][]{Gyulassy08,Edelsbrunner10}, with the advantage of opening the path towards 
a fully hierarchical formalism. This intimate relationship was also recognized by
\cite{Sousbie09}, who used a probabilistic extension of the watershed technique in the latest version
of the skeleton formalism.

\subsection{The Discrete Watershed Transform}
The implementation of the watershed transform in a large variety of scientific applications has to address 
a few important practical issues. A typical characteristic of most scientific images is their discrete nature. 

The discreteness concerns two aspects: the spatial discreteness, ie. the discrete
number of intervals at which the image has been sampled (pixels/voxels), and the discrete intensity levels at which
the image has been sampled.

Image discreteness creates a few complications for an accurate calculation of the 
watershed transform. It renders it difficult to identify the existence and exact 
location of saddle points on the basis of a discretized local neighborhood. 
For the same reason, it is difficult to accurately extract slope lines. 

Several methods for the extraction of critical points have been developed in an attempt to 
alleviate the limitations imposed by the discreteness of images. Among these, the 
\textit{discrete watershed transform} algorithm \citep{Beucher82} represents a simple 
and elegant formalism for identifying the watershed separatrices and can be shown to 
converge to the continuous case \citep{Najman94}.  
The procedure emulates the flooding of valleys or \textit{catchment basins} in a  (discrete) image 
representing a landscape. The points where two or more lakes converge are marked, and the 
algorithm continues until all the pixels in the image have been flooded. At the end of 
the process the image will be \textit{segmented} into individual regions sharing a local minima, 
with the points that were marked as the dividing boundaries between two or more valleys defining 
the {\it watershed transform}.

A major asset of the intensity discretization is that it helps to remove faint features, and therefore also removes 
artefacts without the need of pre- or post-processing. Perhaps the greatest advantage 
is that discrete images allow the use of highly efficient algorithms but in general their use
is limited to image segmentation since they give incorrect topologies.

In the case of images with continuous (floating point) values one retains the option of 
computing the watershed transform directly from the continuous intensity image, in addition 
to the option of discretizising the intensity. On the basis of the continuous image, the watershed 
transform would delineate the topology more accurately than would be feasible on the basis of the 
the discrete-level representation. However, it would involve a substantial increase in computational cost 
and of complexity of the code.

\section{Spine of the Cosmic Web}
\label{sec:cosmicweb}
The analogy between the watershed transform defining the boundary between underdense basins and the 
topology of the cosmic matter distribution is in itself one of the major justifications of the SpineWeb 
method presented in this study. Basic is the connection between the elements that form the Spine of 
the Cosmic Web: walls, filaments, clusters and voids.

\begin{enumerate}
\item[$\bullet$] The Cosmic Web is an interconnected system of dense compact clusters, elongated filaments 
and tenuous sheetlike walls. Visible through the galaxies, gas and dark matter populating these structural 
features, the Cosmic Web theory \citep{Bond96} teaches us that its topological outline was already 
present in the primordial perturbation field out of which all structure arose. 

\item[$\bullet$] All of the elements of the Cosmic Web are interconnected. This is a crucial observation, which can 
be most readily appreciated by studying high resolution N-body simulations \citep[e.g.][]{Springel05}.
Otherwise seemingly isolated objects usually turn out to be connected to less massive structures which 
become visible when assessing the mass distribution at a higher mass resolution. A tantalizing
idea is that the galaxies found at the center of voids lie at the intersection of tenuous 
intra-void dark-matter filaments.\citep{Zitrin09,ParkLee09,Stanonik10}.

\item[$\bullet$] Filaments are suspended between clusters or, dependent on scale, massive halo clumps. 
Their prominence and density may vary substantially, dependent on the mass, distance and alignment of 
the generating dark matter halos. However, the sheer presence of two matter clumps is already sufficient 
for the corresponding tidal force field to guarantee the topological presence of a filamentary bridge. 
Tenuous membranes permeate the space between adjacent filaments, and are part of the large wall which 
defines the boundary between two underdense voids. The wall boundary is outlined by various filaments, 
connecting each other at the cluster nodes. 
\end{enumerate}

Following these observations, the \textit{Cosmic Spine} is defined as the topological network of nodes, filaments 
and sheets along which the cosmic matter distribution on large Megaparsec scales has assembled (see 
fig.~\ref{fig:density_landscape}). 

\section{The Spineweb Procedure}
\label{sec:spineweb}
The key aspect of the SpineWeb procedure is that it exploits the intrinsic topological information contained 
in the {\it Watershed Transform} to delineate the Cosmic Spine. For the computation of the Cosmic Spine by means 
of the Watershed Transform it is necessary to address a few issues of practical importance. 

\subsection{Density field}

\label{sec:dtfe}
In our current implementations of the SpineWeb procedure, we apply the DTFE method to 
reconstruct the density field from the spatial particle distribution. 

The DTFE procedure produces a self-adaptive volume-filling density field on the basis of 
the Delaunay tessellation of the point distribution\citep{Schaap00,Schaap07}. DTFE 
density (and velocity) fields have been found to optimally trace a hierarchical matter distribution at 
any resolution level represented by the point sample, while at the same time resolving the local anisotropies 
in the matter distribution. This high level of sensitivity to the topology of the matter distribution, 
makes DTFE ideally suited for the SpineWeb procedure \citep[for an extensive description of the DTFE 
procedure, see][]{WeySchaap09}. 

Given a spatial distribution of points, DTFE is based on the assumption that the density at the position of 
each point is proportional to the inverse of the total volume of the adjacent Delaunay tetrahedra, ie. 
to the volume of its contiguous Voronoi cell. Subsequently, the density field values at any location 
throughout the sample volume is determined by means of linear interpolation within the Delaunay 
tetrahedra of the corresponding Delaunay tessellation. Because a singular density
determination at the central location of a voxel tends to introduce aliasing artefacts at high densities, we
follow a slightly elaborate procedure. The density at each voxel of the ``image'' grid is
determined on the basis of the DTFE density field sampled on a subgrid with a three times higher
resolution and the density at each gridpoint set equal to the mean of the DTFE values at the 27
subgrid locations within the corresponding voxel.

In the applications described in this study, we compute the density field values at the voxels of 
a regular cubic grid. We use a fast and efficient implementation of the DTFE algorithm based on the publicly available
CGAL library\footnote{www.cgal.org}.

\subsection{Watershed Implementation}
The discrete watershed transform code we use in the SpineWeb procedure is an 
adaptation of the immersion and the topographical distance algorithms for floating point intensity values 
\citep[see][for a review]{Roerdink00}. The code assumes a density map which is sampled 
on a regular grid. Our \verb=C= code \footnote{The code will be publicly 
released in an upcoming article. In the meantime it can be provided upon request.} computes the 
watershed transform from a double-precision $512^3$ grid in just a couple of minutes on a regular 
linux workstation. 

In a first step, the code starts by finding and labeling the local minima in the density map, by 
identifying the voxels with the lowest density value among all their 26 neighbors. These local minima 
are the seeds of the void valleys to be identified by the watershed transform. 

In a second step, we follow the topographical distance algorithm in order to obtain a fast  
segmentation of the space into locally connected underdense regions. For each voxel we identify 
the voxel among its 26 neighbors which has the lowest density. The maximum gradient paths are 
traced by iteratively connecting the voxels to their lowest density neighbor until the path 
reaches a local minimum. Subsequently, we assign the label of the corresponding minimum to the path. 

In the third step we extract the watershed transform itself, ie. the boundaries between the void regions. 
The pixels in the watershed transform are identified by means of a \textit{local immersion algorithm}. 
First, we identify all the voxels that lie at the boundaries between two or more regions. Subsequently, 
this subset of voxels is sorted in density and the standard immersion algorithm is applied. By following 
this two-step procedure, we avoid what would be the most expensive component of the algorithm. Instead of 
having to sort the complete density field, the sorting evaluations are restricted to the points in 
the watershed boundaries, a minor fraction of the complete volume. 

In a final step, each of the pixels in the watershed boundary is assigned a morphological label, 
following its identification as \textit{void}, \textit{wall} or \textit{filament} element, according 
to the criterion expressed in equation~\ref{eq:neighbour_morphology} and as illustrated in 
fig.~\ref{fig:simple_neighbourhoods}.

While the watershed transform provides us with a highly efficient means of segmenting space into 
topologically well-defined elements, in general the watershed algorithm tends to overdo the 
segmentation, creating too many regions and identifying small noise in the image rather than 
real features \citep[see][]{Edelsbrunner10}. We circumvent this circumstance by preprocessing the 
density or distance field via a Gaussian smoothing of $\sigma=2$ voxels.

\subsection{from Watershed to SpineWeb}
From the analogy between the Cosmic Web and the watershed transform one can define, on the basis of the 
discrete watershed transform of a cosmic density field, a set of unique criteria to identify 
voids, walls and filaments.

The criteria are based on the properties of the local neighborhood of all the points 
that comprise the discrete watershed transform. Instead of computing at any given point  
the local eigenvalues of the Hessian of the density field, one may simply resort to the entirely 
equivalent evaluation of the identity of the surrounding 26 neighbor pixels (for the three-dimensional 
situation). By counting the number $\mathcal{N}_{\textrm{\tiny{voids}}}$ of adjacent watershed basins (voids) 
amongst these, it is straightforward to discriminate between voxels which belong to a void, a wall or a 
filament by means of the following set of rules:   
\begin{equation}
\mathcal{N}_{\textrm{\tiny{voids}}}\qquad \left\{
 \begin{array}{rl}
     \quad =  \;\;\; 1, \qquad & \text{void} \\
     \quad =  \;\;\; 2, \qquad & \text{wall} \\
     \quad \geq \;\;\; 3, \qquad & \text{spine} \\
     \quad \qquad & \text{(filament + clusters)} \\
 \end{array} \right.
\label{eq:neighbour_morphology}
\end{equation}
\noindent Note that in the present implementation we do not discriminate between filament or 
cluster node, and instead consider them to be part of the same spinal structure. In a future 
implementation we will include a density maximum criterion which would allow us to find the 
cluster nodes amongst the spine voxels. In addition we can easily identify regions in clusters
by using one of the common halo finders available like friends of friends and isolate them from the
identified filaments.

Our Spineweb technique exploits a purely local criterion for identifying the 
morphological nature of boundary pixels in the density field: it utilizes the full 
geometric structure of the watershed transform to limit the evaluation to the direct 
neighbours of each point. This differs from the implementation followed by \cite{Sousbie09} who
included a probability propagation scheme in the watershed flooding procedure from
where the different elements of their skeleton were determined by finding their 
intersections between the corresponding peak and void patches.

\begin{figure}[!h]
  \centering
  \vskip 0.5truecm
  \includegraphics[width=0.5\textwidth,angle=0.0]{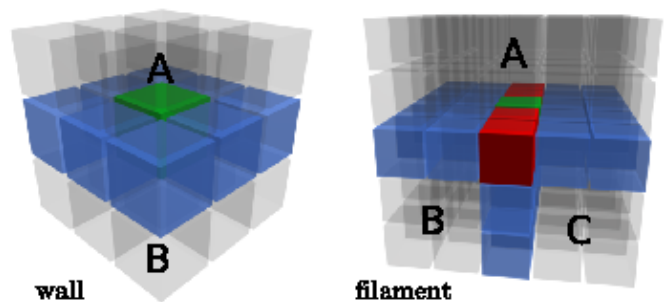}
    \caption{Local neighborhood around a voxel (green) inside a wall (left) and filament (right).            
	            The blue voxels indicate walls and the red voxels filaments. The light gray cubes here represent
	            voxels inside voids. The voxel inside a wall has two
	            adjacent voids inside its neighborhood (A and B) while the voxel inside a filament has in this case
	             three adjacent voids (A, B and C).}
  \label{fig:simple_neighbourhoods}
\end{figure} 

The above criterion is a purely and solely a topological one. By definition, walls are the 
regions between two adjacent voids. Filaments are to be found at the intersection of three 
watershed basins, at the intersection of 3 walls (see fig. \ref{fig:simple_neighbourhoods}). The success of these criteria can be 
appreciated from the 3-D surface maps in fig.~\ref{fig:spine_filas_walls_surfaces} and the 
comparison between density and spine maps in fig.~\ref{fig:spine_densi_filament_walls_ghost}.

\subsection{Image Grid Representation}
While a regular grid facilitates the computation of the watershed transform and the 
subsequent topological identification of the various boundary pixels (see fig.~\ref{fig:simple_neighbourhoods}), 
its simplicity may also involve a few possibly artefacts. 

The first artefact relates to the discrete nature of the voxels in the density field. As  a
results, the filaments and walls have an artificial thickness - even if they would be 
infinitesimally thin - which makes the look \textit{pixelated} or \textit{jagged}. A 
particularly good illustration of this is the Spine obtained for the Voronoi clustering 
model in fig.~\ref{fig:voronoi_results} (panel c) where we explicitly render individual 
voxels as cubes. In the asymptotic limit of infinitesimally thin voxels 
the discrete watershed will converge to the continuous case. 

The second artefact concerns the anisotropic nature of the local neighborhood 
of each voxel: Each voxel has 6 neighboring voxels at a distance $d=1$ (in voxel units), 
16 neighbors at $d=\sqrt{2}$ and 8 neighbors at $d=\sqrt{3}$. 
it also involves an angular neighbor distribution deviating   
substantially from angular isotropy.  A possible alternative would be to limit the 
neighborhood evaluation to the 6 most direct neighbors. However, the poor sampling 
might lead to a considerable risk of missing important topological information. 
For two-dimensional images the solution would be more straightforward. The use 
of a hexagonal grid would involve equal distance for all neighbor pairs and a 
perfectly uniform angular distribution. Unfortunately, an equivalent perfect grid 
for the three-dimensional situation does not exist. However, the use of Centroid 
Voronoi Tessellations (CVT, \cite{Du99}) would certainly help to alleviate the main 
artefacts.

\subsection{Galaxy Spine Assignment}
Physically, filaments and walls are not infinitesimally thin structures. To identify the particles 
or galaxies attached to them, we therefore need to the define a (natural) thickness which encloses 
these objects. 

In the applications described below, we account for this by applying the \textit{dilation} 
morphological operator to the voxels labeled as filament and wall. The process increases the thickness of 
filaments and walls by one voxel and this procedure can be performed iteratively to further increase 
the thickness. 
The dilation operator was applied first to voxels labeled as wall 
and subsequently to pixels labeled as filaments following the number of degrees of freedom in the local 
variation of the density field, i.e. first walls and subsequently filaments \citep{Aragon07b}.
In our particular case a single iteration with a $3\times3\times3$ kernel provides a good result without 
excessively fattening the structures. 

\section{SpineWeb Test:\\ \ \ \ \ Voronoi Clustering models}
\label{sec:voronoi}
To test and quantify in an objective way the identification of walls and filaments 
with the SpineWeb procedure we apply it to a few realizations of Voronoi 
Clustering models of the large scale matter distribution \citep{Weygaert89,Weygaert02,Weygaert2010}. 

\begin{figure*}[!ht]
  \centering
  \includegraphics[width=0.98\textwidth,angle=0.0]{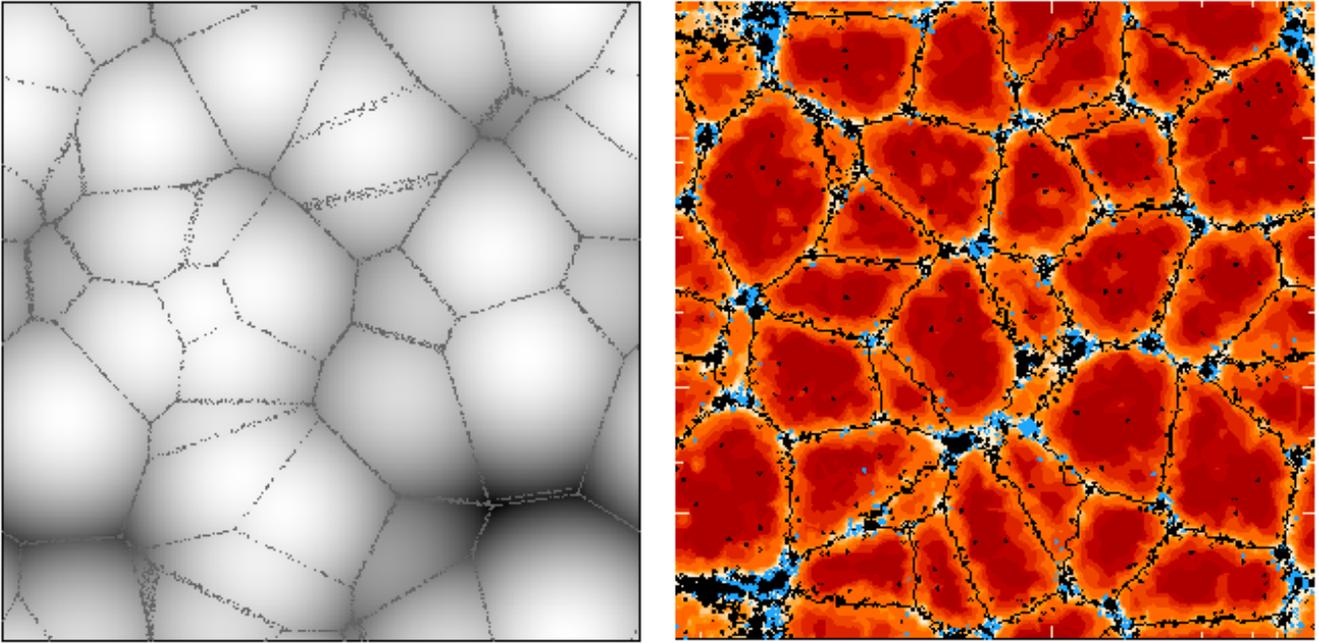}
  \caption{Left: slice of the distance field computed from the Voronoi seeds (background). The particles located inside
           filaments, walls and nodes are also shown. Right: Voronoi model with density field reconstruction. A slice 
           across the intensity of the density field is shown in orange scale in the background. The black lines 
           correspond to the watershed transform and the particles are indicated by small diamonds. Blue particles are 
           misclassifications (in all the categories) black particles are correct classified particles.}
  \label{fig:voronoi_distance_field}
\end{figure*}

The Voronoi clustering models are heuristic models for cellular spatial
patterns which use the geometric (and convex) structure of the Voronoi tessellation 
\citep{Voronoi1908,Okabe2000} to emulate the cosmic matter distribution. They offer 
flexible templates for cellular patterns and are easy to tune towards a specific spatial 
cellular morphology. This makes them very suited for studying clustering properties of 
nontrivial geometric spatial patterns. Because the location, geometry and identity of the 
various spatial components in Voronoi models are known precisely, they are ideal as testbeds 
for the SpineWeb procedure. Unless otherwise specified, the seeds of the 
tessellation usually involve a set of Poisson distributed pints. 

The Voronoi models use the corresponding tessellation for defining the structural
frame around which matter will gradually assemble during the formation and growth of 
cosmic structure. Points are distributed within this framework by assigning them to 
one of the four distinct structural components of a Voronoi tessellation: 
\begin{enumerate}
\item[$\bullet$] \emph{Void}:\\regions located in the interior of Voronoi cells.
\item[$\bullet$] \emph{Wall}:\\regions within and around the Voronoi cell faces.
\item[$\bullet$] \emph{Filament}:\\regions within and around the Voronoi cell edges.
\item[$\bullet$] \emph{Clusters}:\\regions within and around the Voronoi cell vertices.
\end{enumerate}
What is usually described as a flattened ``supercluster'' consists of an assembly of various 
connecting walls in the Voronoi foam, while elongated ``superclusters'' of ``filaments'' 
usually include a few coupled edges. Vertices are the most outstanding structural 
elements, corresponding to the very dense compact nodes within the cosmic web where 
one finds the rich clusters of galaxies. 

Among a variety of possible Voronoi clustering realizations, two distinct yet 
complementary classes of models are the most frequently used ones, the structurally 
rigid {\it Voronoi Element Models} and the evolving {\it kinematic Voronoi models} 
\citep[see e.g.][for an extensive description]{Weygaert2010}. Here we use one 
Voronoi Element Model, a composite of all four distinct components, and one realization 
of a Voronoi kinematic model. 

In the case of the Voronoi Element model, the walls, edges and vertices are infinitely thin, 
yielding a pure geometric discrete realization of the underlying Voronoi tessellation. The 
Voronoi kinematic model represents a more realistic situation in which the various 
structures are assigned a finite width. We first analyze the topologically cleaner 
configuration of the Voronoi Element model, to assess the performance of SpineWeb under 
optimal conditions. Subsequently, we investigate whether in how far it can sustain 
this performance under less optimal circumstances. 

\subsection{Voronoi Element model}
Based on a uniform distribution of $M$ cell seeds in a (periodic) box of size L, 
we start with a uniform distribution of N particles throughout a (periodic) box 
of size L. The particles are distributed within the tessellation by projecting them - with 
respect to the seed of the cell in which they are originally located - onto the walls, edges 
or vertices surrounding their Voronoi cell. As a result each of the walls, filaments and 
vertices have a different density, although the density remains uniform within each of the 
individual elements. 

The Voronoi Element models we used for our test consisted of a set of particles distributed in a
box of size L with $10^6$ particles from which $70\%$ resides in the walls, $25\%$ 
in the filaments and $5\%$ in the clusters. We chose a realization completely 
devoid of particles in the interior of voids. This distribution of particles was chosen 
in order to obtain a more uniform sampling over the three morphologies compared
to a more realistic distribution.

\subsubsection{Voronoi Distance Field}
For the purpose of this test, instead of basing the SpineWeb procedure on the reconstructed (and noisy) density field
we use the knowledge of the underlying tessellation to define a clean \textit{distance field}. 

The main idea behind the SpineWeb method, the identification of morphological structures on the 
basis of their topology, does not depend on practical details of the density field 
determination from a dataset of observed galaxy locations or computer simulations.  In this respect, 
it is important to realize that the validity of the SpineWeb procedure can be tested and assessed 
on the basis of the topological structure of any field that is topologically equivalent to the 
density field of the Cosmic Web. This indeed is true for the distance field, and any generic field marked 
by a monotonic increasing value from a field minimum towards its watershed transform. 

In the particular implementation described in this section, the distance field is defined as the Euclidean 
distance from each particle to its closest Voronoi seed. 
Regions close to the cell centers have low values while regions in the planes and edges of the cell have 
large values, with the value gradually increasing along the direction from the Voronoi cell centers towards 
the projected location on the walls. Within the walls, the highest density is reached at the surrounding edges, 
ultimately peaking at the vertices. In the resulting distance field, small cells correspond to low field 
values, while the larger cells yield higher field values, particularly near their boundaries.
Following this definition, the distance field emulates the range of densities encountered in the Cosmic Web. 
The finite volumes of Voronoi cells in periodic tessellations assures a 
convergence of the distance field on the boundaries of data volume. 

Various definitions of distance fields might be used, largely dictated by the specific questions at hand. An 
example of one such possibility would be to normalize the distance field on the basis of the distance to of the 
point to its second closest Voronoi nucleus, or the distance of its projected location on the corresponding Voronoi 
wall. The resulting distance field would reach a value of unity at the walls of the Voronoi cells. However, we opted 
to use the distance field with no normalization since it gives a better representation of the large dynamical range 
of densities encountered in the Cosmic Web. 

\subsubsection{Distance Field Realization}
For the Voronoi Element test model, we determined the distance field on a cubic $512^3$ grid. For each of the 
pixels in the grid we identify the closest nucleus, among the set of $M$ generating nuclei. The pixel is assigned 
the value of its \textit{Euclidean distance}. 

The resulting field is shown in fig.~\ref{fig:voronoi_distance_field}. It depicts the 
distance field itself, in a planar section through the 3-D box, by means of a greyscale map. The corresponding 
particles located in the filaments, walls and nodes, within a narrow strip around the sectional plane, are 
superimposed on the image. 

\begin{figure*}[!ht]
  \centering
  \includegraphics[width=0.99\textwidth,angle=0.0]{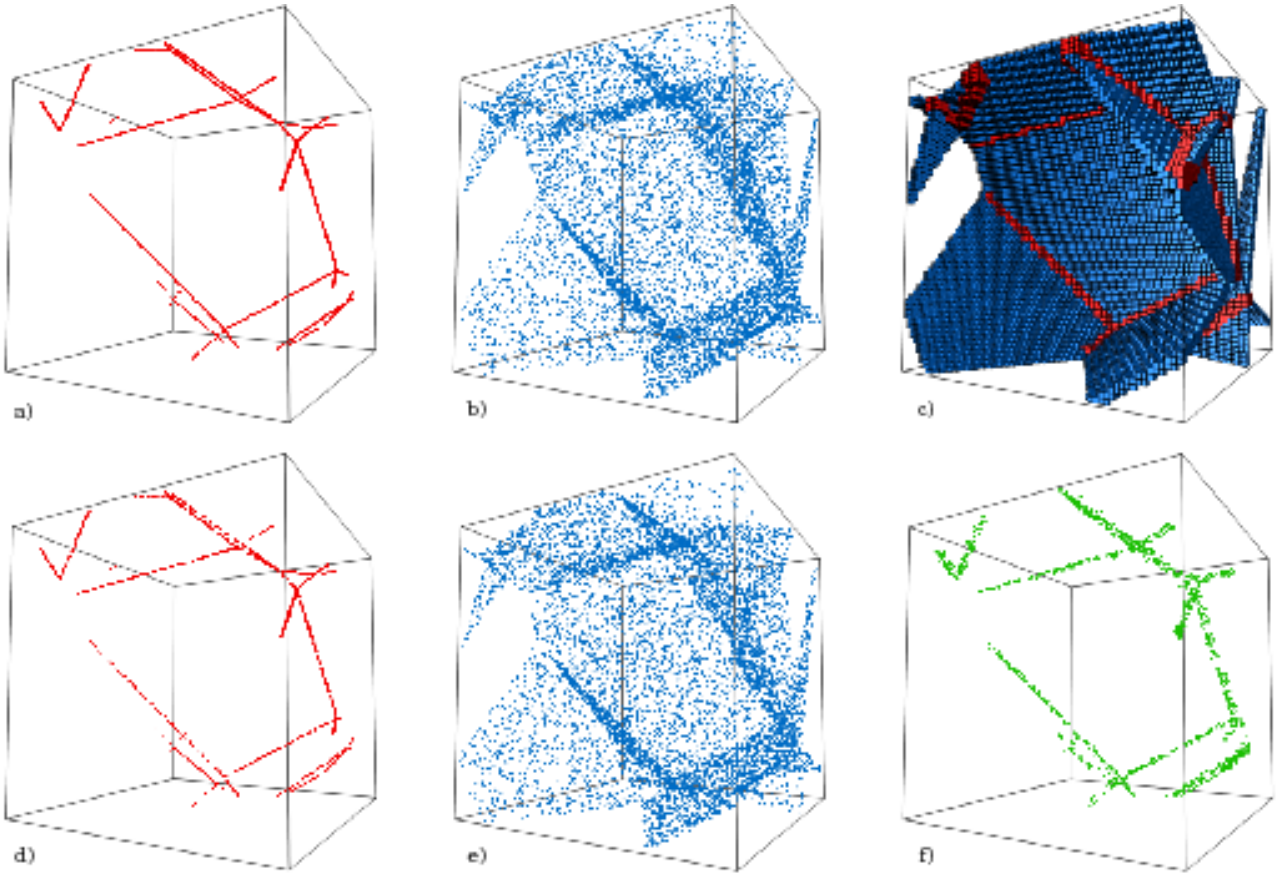}
  \caption{The SpineWeb method applied to a Voronoi distribution.
	\textbf{a)} original particles lying at the edges of the Voronoi cells (filaments).
	\textbf{b)} original particles lying at the faces of the Voronoi cells (walls).
	\textbf{c)} Pixels inside filaments (red) and walls (blue) identified with the SpineWeb method.
	\textbf{d)} recovered particles lying at the edges of the Voronoi cells (filaments).
	\textbf{e)} recovered particles lying at the faces of the Voronoi cells (walls).
	\textbf{f)} particles erroneously identified as particles in filaments.
	The box shown here contains $1/64$ of the original box volume.}
  \label{fig:voronoi_results}
\vskip 0.5truecm
\end{figure*}

\subsection{Voronoi Kinematic Model}
A nearly equivalent Voronoi clustering model realization is a Voronoi Kinematic Model. Here initially 
randomly distributed particles move away from their expansion nucleus - ie. the closest nucleus in whose 
Voronoi cell they are located - by a universal expansion rate \citep[see e.g.]{Weygaert02,Weygaert2010}. 

When particles reach the wall shared between their expansion center and the second closest nucleus, their 
motion is constrained to their path within the wall. This continues until they reach one of the edges 
delimiting the wall, upon which they proceed along the edge towards their ultimate location, that of 
one of the vertices at the ends of the edge. 

The model simulation box has a length of 141h$^{-1}$Mpc, in which we find 180 Voronoi cells. In total, the 
box contains $N=2097152$ particles. Originally distributed randomly throughout the box, we move them 
until 16.5$\%$  of the galaxies reside in the walls, 28.7$\%$ in the filaments and 51.3$\%$ at the cluster 
nodes. Unlike the Voronoi Element Model described in the previous section, the voids remain populated with 
a diluted random distribution of 3.5$\%$ of all void galaxies.

The different morphological structures in this Voronoi kinematic model are also assumed to have a finite physical 
width. Particles within the walls, edges and vertices are assumed to have a Gaussian distribution perpendicular to 
these structures. The width for each of these morphologies is set to $\sigma = 1.0$h$^{-1}$Mpc. Even though 
the topological properties of the pure Voronoi Element Model and this Voronoi kinematic model are practically 
equivalent, the finite width and more organic development of the Voronoi kinematic models represent 
a spatial density field which more closely emulates that encountered in galaxy redshift surveys and in 
N-body simulations of structure formation. 

\subsubsection{Voronoi Model Density field}
The SpineWeb performance test for the Voronoi kinematic model is based on the density field of 
the particle distribution. We use the DTFE method to reconstruct the density field from the 
point distribution on a 256$^3$ cubic grid within the simulation box.

Figure~\ref{fig:voronoi_distance_field}, righthand panel, shows a 2-D section through the reconstructed 
density field, which is represented by a color map. The low-density patches and high-density structures clearly 
outline the interior region of voids, while the high-density regions correspond to the walls and 
filaments in the overall spatial pattern. 

\subsection{SpineWeb identification}
Following the determination of the distance field for the Voronoi Element Model, and the 
density field for the Voronoi kinematic model, the SpineWeb procedure proceeds to identify 
the Spine of the particle distribution. Following the identification, we assess the 
fraction of false and real spine detections in both Voronoi  models. 

\subsubsection{Detection Rates: definition}
Quantitatively, we assess the detection rate of the SpineWeb procedure by determining the ratio of 
real and false detections of filament and wall particles. 

The detection rate $R_{\textrm{\tiny{real}}}$ is defined as the fraction of original wall or filament particles that are 
also identified as such by the SpineWeb technique, 
\begin{equation}
R_{\textrm{\tiny{real}}}\,=\,\frac {N_{\textrm{\tiny{real}}} }{ N_{\textrm{\tiny{original}}}}\,.
\label{eq:realdet}
\end{equation}
\noindent In this equation, $N_{\textrm{\tiny{real}}}$ is the number of wall or filament particles that 
have also been identified as such, and $N_{\textrm{\tiny{original}}}$ is the total number of particles 
that in the Voronoi model physically belong to a wall or filament. Along the same lines, we measure the 
filament and wall contamination rate $R_{\textrm{\tiny{false}}}$ following the definition, 
\begin{equation}
R_{\textrm{\tiny{false}}}\,=\,\frac {N_{\textrm{\tiny{false}}} }{ N_{\textrm{\tiny{original}}}}\,.
\label{eq:falsedet}
\end{equation}
\noindent 
\noindent $ N_{\textrm{\tiny{false}}} $ is the number of particles recognized as wall or filament 
particle, but in reality having a different morphological identity. 

The results for the the detection and contamination rates of walls and filaments in both the 
Voronoi Element model as well as the Voronoi kinematic model are listed in table~\ref{tab:recovered_rate}.

\begin{table}[!h]
  \caption{Recovered particles per morphology}
  \label{tab:recovered_rate}
  \begin{center}
    \leavevmode
    \begin{tabular}{lll} \hline \hline              
Structure        & $ R_{\textrm{\tiny{real}}} $  & $ R_{\textrm{\tiny{false}}} $ \\
\hline 
Walls$_{\;\textrm{\tiny{DIS}}}$     &  0.93 & 0.15 \\
Spine$_{\;\textrm{\tiny{DIS}}}$     &  0.91 & 0.03 \\
\hline
Walls$_{\;\textrm{\tiny{DEN}}}$     &  0.76 & 0.32 \\
Spine$_{\;\textrm{\tiny{DEN}}}$     &  0.87 & 0.13 \\
\hline
    \end{tabular}
\caption{Ratio of real and false recovery rates per morphology. The top half corresponds to the
	results from the distance field (dis), while the bottom half concerns the results for the 
        DTFE reconstructed density field (den). For definition of $ R_{\textrm{\tiny{real}}}$ and 
        $R_{\textrm{\tiny{false}}}$ see eqn.~\ref{eq:realdet} and ~\ref{eq:falsedet}}
  \end{center}
\vskip 0.5truecm
\end{table}

\subsubsection{Spine of the Voronoi Element Model}
In order to remove small-scale spurious variations, the distance field is smoothed with a 
Gaussian filter of $\sigma = 2$ voxels. Of the smoothed distance field we compute the 
watershed transform. Following this, the Spine is determined. 

The SpineWeb results for the Voronoi Element Models are illustrated in figure~\ref{fig:voronoi_results}. 
Visual inspection of the figure provides a good impression of the virtues and performance 
of the procedure. Comparison between panels~(a) and ~(d) shows that the genuine filament particles 
in the Voronoi model (a) are identified with a convincing accuracy by the SpineWeb procedure (d). 
The same is true for the successful identification of the Voronoi wall particles (panel b) and the 
SpineWeb identified wall particles (panel e). Interestingly, the particles that SpineWeb erroneously 
identify as filament particles while in fact they are wall particles, shown in panel~(f), clearly 
delineate the original filamentary web. It is most most likely a result of the discrete resolution 
of the distance field grid. 

The SpineWeb reconstructed spine, shown in panel~(c) of fig.~\ref{fig:voronoi_results}, allows 
a clear and transparent assessment of the topological structure of the Voronoi web. It shows the 
identified wall voxels by means of blue blocks, while the filamentary voxels are indicated as 
red blocks. The walls form a continuous network of connecting surfaces, with filaments delineating 
the intersections between the walls. 

From table~\ref{tab:recovered_rate} we can see that $93\%$ of particles in walls and  $91\%$ of particles 
in filaments are also identified as such, while the contamination rate of walls and filaments are 
$15\%$ and $3\%$. The false identities can usually be ascribed to the jagged nature of the voxels 
(see panel~(c), fig.~\ref{fig:voronoi_distance_field}).

The simple and idealized example of the Voronoi Element model shows the intrinsic potential and 
power of the SpineWeb procedure. Solely on the basis of the topology of the matter distribution, 
and independent of a density threshold or any other arbitrary parameter, it manages to 
determine its correct morphological segmentation.

\begin{figure*}[!ht]
  \centering
  \includegraphics[width=0.49\textwidth,angle=0.0]{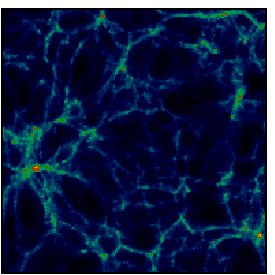}
  \includegraphics[width=0.49\textwidth,angle=0.0]{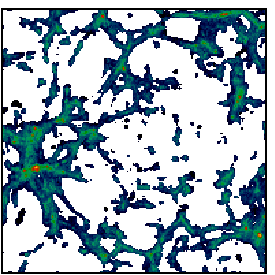}
    \caption{Volume rendering of a thick slice of the density field in a $\Lambda$CDM simulation in a box 
     of 200$h^{-1}$Mpc. Left: the full DTFE reconstructed density field. Right: the density field \textit{inside} 
     an isosurface at $\delta=1$.}
  \label{fig:density_isosurface}
\end{figure*} 

\subsubsection{Spine of the Voronoi Kinematic Model}
The situation for the Voronoi Kinematic Model, marked by a more noisy particle distribution and a less 
idealized density field reconstruction, is less straightforward yet more representative for realistic 
circumstances. 

Fig.~\ref{fig:voronoi_distance_field} (righthand panel) shows, superimposed on the density field 
color map, the Voronoi particle distribution as well as the watershed transform. The latter is 
shown by means of the black lines. Particles identified as wall or filament particles are indicated 
by small diamonds. The watershed transform is able to identify most of the voids and their boundaries. The image 
shows that the SpineWeb procedure managed to closely follow the location of the edges and faces 
in the original Voronoi tessellation. 

Some minor deviations are detected at small scales, the result of Poisson noise in combination with the 
discrete nature of the sampled density field. By construction, our filaments and walls are only one 
voxel thick (in this Voronoi model realization this is equal to $\sim 1 h^{-1}\textrm{Mpc}$). As a result, we 
may miss the particles inside a given structure because of small-scale 
variations in the density field translating into variations in the watershed transform. This effect 
can be appreciated from the miss-classified (blue) particles in fig.~\ref{fig:voronoi_distance_field}, 
mostly clustered around the boundaries between filaments and filaments. 

To determine the detection and contamination rate of the SpineWeb calculation, we compare the 
identity of the Voronoi wall and filament particles, which are a priori known from the 
model generation, with that of the classification on the basis of the reconstructed 
watershed segmentation. To this end, we assign the particles within a radius of 
$2\sigma\sim 2h^{-1} \textrm{Mpc}$ from a wall or a filament in the spine (watershed) 
segmentation to that particular structure: a particle is identified with a wall when 
it it lies within a $2\sigma$ distance from two different watershed cells. Although the 
detection rate results are less forthcoming than for the pure Voronoi element models, 
they remain convincing (see lower half of table~\ref{tab:recovered_rate}. We find a detection 
rate of $76\%$ for wall particles, as opposed to a contamination rate of $\sim 32\%$. Filaments 
are better recognizable, which may be understood from the $87\%$ detection rate of filament 
particles, as opposed to a mere $13\%$ misclassification rate. 

\vskip 0.5truecm
\section{Single-scale $\Lambda$CDM Spine}
\label{sec:lcdm}
To test the SpineWeb method in a more realistic and challenging situation we have applied  
it to a cosmological N-body simulation. It concerns a $\Lambda$CDM universe simulation inside a box 
of 200 $h^{-1}$ Mpc, restricted to the dark matter particles. Initial conditions were generated on 
a $512^3$ grid with $\Omega_m = 0.3$, $\Omega_{\Lambda} = 0.7$, $\sigma_8 = 0.9$ and $h = 0.73$. 
For the primordial perturbations, we use the transfer function of \citet{Bardeen86} with a  
shape parameter $\Gamma=0.21$ following the definition of \citet{Sugiyama95}. After having set up 
the initial conditions, we follow the subsequent gravitational evolution to the present time using 
the public N-body code Gadget2 \citep{Springel05}.

From the same initial conditions, we also generated additional lower-resolution versions of 
$256^3$, $128^3$ and $64^3$ particles were generated, following the ``averaging'' prescription described in 
\citep{Klypin01}. For the single-scale analysis in this paper, which focuses on the largest filaments and walls 
in the particle distribution, it is sufficient to analyze the low-resolution $64^3$ dataset. The lower 
resolution corresponds to a cut-off scale of $\sim 3h^{-1}$Mpc in the initial conditions, sufficient 
for the analysis focusing on the large scale structure. The higher resolution datasets are used for 
visualization purposes. 

\subsection{Density field morphology}
\label{sec:densmorph}
From the final particle distribution we compute the density field on a cubic grid of 512 voxels per 
dimension using the DTFE method (see sect.~\ref{sec:dtfe}). Figure~\ref{fig:density_isosurface} 
(lefthand panel) depicts a volume rendering of the density field in a thick slice through the simulation box. 
It shows that DTFE manages to follow the intricacies of the weblike structures in great detail, over a range of scales: 
it reproduces the correct geometry of the various features. 

\bigskip
An interesting example of structural complexities in the displayed region is the cluster at the lefthand side 
of the slice. A full 3D visualization of the system shows that the filaments entering the cluster define 
several semi-planar structures, all sharing the cluster as their common node. Lower isodensity contours 
reveal even more of the tenuous walls, even though at such low density levels we need to take into 
account that the image gets easily confused by spurious interloping features. 

A frequently used approach for delineating structural features such as filaments and walls is 
to assign a specific density range to each morphology. Filaments or walls are singled out 
by selecting the regions that have a density within the corresponding density range. In the righthand panel 
of fig.~\ref{fig:density_isosurface} we show the density field inside an isodensity surface at $\delta=1$, 
for clarity superimposed on top of a white background. Because a density value $\delta=1$ is roughly comparable to 
typical values encountered in filaments and walls \citep[see e.g.][]{Aragon07}, these contours roughly 
define the boundaries of the filaments and the clusters embedded, along with the tenuous walls suspended 
between the filaments. 

\begin{figure*}[!ht]
  \centering
  \includegraphics[width=0.75\textwidth,angle=0.0]{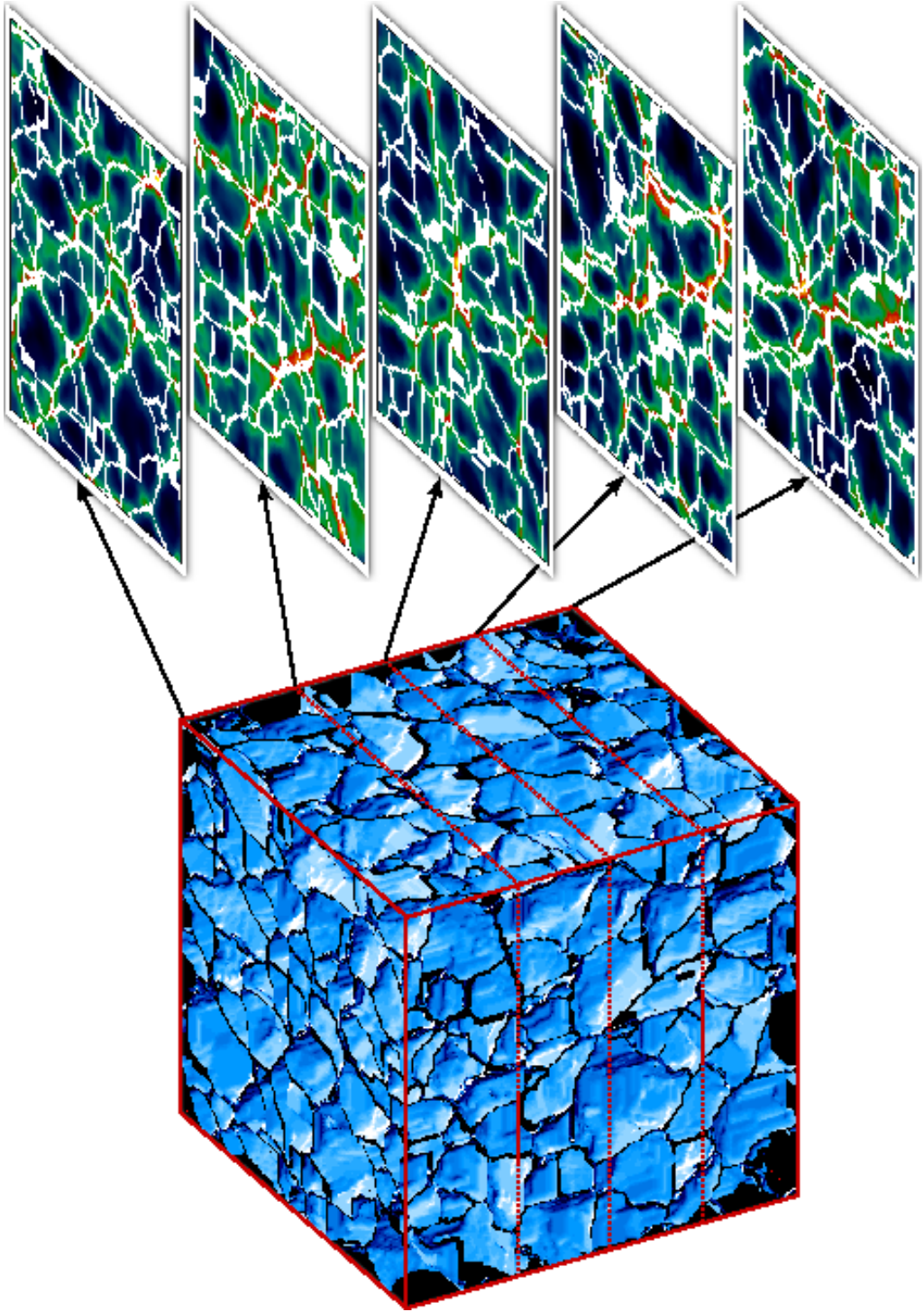}
    \caption{Three-dimensional watershed transform of the density field. The cube shows the pixels that compose 
      the watershed transform, from which the Spine is extracted. Several slices cut across the 
      simulation box show the watershed (white lines) delineating the density field (blue-green background). 
      The three dimensional nature of the watershed network is evident.}
  \label{fig:watershed_slices}
\end{figure*} 

Although the isodensity surfaces provide good insight into the overall distribution of matter, one immediate 
observation is that the attempted pure density selection of filaments is not very successful. As was pointed out 
by \cite{Aragon07} filaments and walls are characterized by a rather broad range of densities 
\citep[also see e.g.][]{Hahn07a}. The broad cluster, filament, wall and field density ranges are 
also mutually overlapping over a sizeable density range \citep[also see][]{Aragon10c}. 

An immediate repercussion of the overlapping density ranges is that is almost impossible to decide purely on 
the basis of a density criterion whether a certain location belongs to a cluster node,filament or wall. This 
may be directly appreciated from the truncated density map in fig.~\ref{fig:density_isosurface}. Although the image 
shows a substantial degree of filamentary structure, comparison with the full density field shows that it  
discards the pattern of lower density filaments. Also, it does not manage to disentangle the highly concentrated 
agglomeration of filaments near the massive cluster at the central lefthand side of the box. Moreover, throughout the whole 
volume it is rather difficult to see which locations would belong to a filament and which ones to a wall. 

\subsection{Cosmic Spine and Cellular Morphology}
Following the computation of the density field, we compute its watershed transform. 
The resulting segmentation of the density field into its watershed basins is illustrated in 
figure~\ref{fig:watershed_slices}. 

The watershed basins are to be identified with the void regions in the cosmic matter distribution. 
To get a better idea of its spatial structure and the connections between the various structural 
components, we slice through the watershed field at regular intervals along the $x$-axis. This 
yields a sample of $yz$-slices through the simulation box. Figure~\ref{fig:watershed_slices} shows 
a sequence of five consecutive $yz$-slices, with the watershed segmentation shown as white lines
superimposed on top of the density field (blue-green level) map.  It is straightforward to appreciate 
the correspondence between the watershed segmentation lines and the underlying density field. 
While the segmented cube in figure~\ref{fig:watershed_slices} emphasizes the characteristic 
cellular nature of the Cosmic Web, the $yz$ slices reveal the close relationship between voids, 
walls and filaments.  

The 2-D $yz$ slices show the strong correlation between the density field and the watershed transform.  
The watershed lines trace the high-density ridges and regions in the density field, occasionally 
bridging their lower-density connecting parts. On a more global scale, we also notice that the 
higher density regions contain a higher number of distinct and smaller cells than the more moderate 
or underdense areas. This translates into a more complex local network of filaments and 
walls. The opposite effect occurs in underdense regions. These are mainly characterized by large 
symmetrical voids, surrounded by relatively simple wall-filament environments.
 
Cosmic voids are immediately recognized as large empty cells in the watershed transform. 
The walls in the cosmic matter distribution are visible as the boundaries between two 
adjacent watershed cells, while the filaments are found at the intersection of these walls. 
The considerable variety of sizes and shapes of voids is most readily visible in the 
pattern of watershed lines in the $yz$-slices. Even though we know that on 
stereological grounds, lower-dimensional sections tend to exaggerate the size 
distribution of the full 3-D distribution, the comparison with the void basins in the 3-D box 
does confirm the impression of the diversity of the void population. It also underlines the 
significance of topological SpineWeb analysis: the void distribution is a direct reflection 
of the complexity of the dynamical processes which are forming and shaping the voids 
\citep{Sheth04,Platen08}. 

\begin{figure*}[!ht]
  \centering
\vskip 0.5truecm
  \includegraphics[width=0.90\textwidth,angle=0.0]{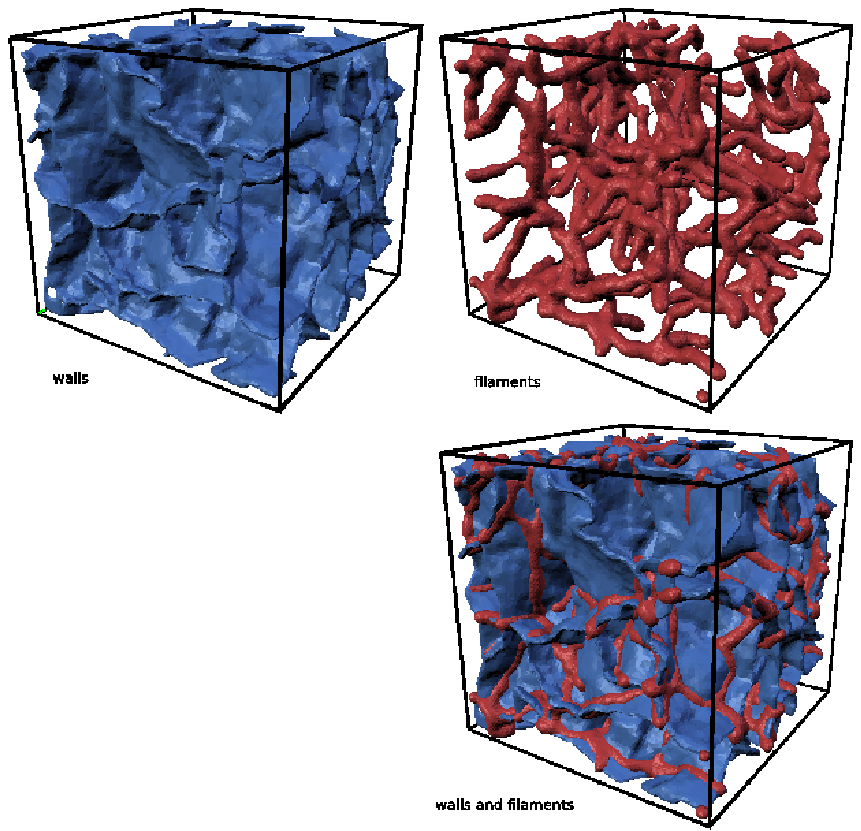}
    \caption{Surfaces enclosing the voxels which are identified as belonging to walls (blue, top left) and 
        filaments (red, top right) within a cubic region of $50 h^{-1}$ Mpc. The bottom frame shows how in 
        the same region both morphological components are connected and intertwined. The latter forms 
        a nice illustration of the intimate relationship between filaments and walls. For visualization purposes 
        the surfaces are smoothed with a Gaussian kernel of $\sigma=2$ voxels.}
  \label{fig:spine_filas_walls_surfaces}
  \includegraphics[width=0.90\textwidth,angle=0.0]{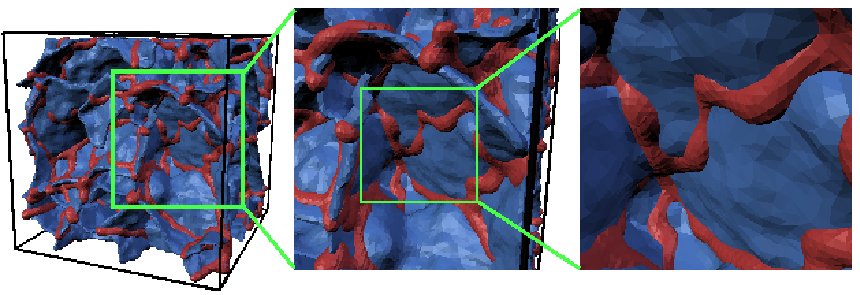}
    \caption{Zoom-in onto the cosmic spine in a subregion of the $50 h^{-1}$ Mpc, highlighting the 
        intricate connections between wall surfaces (blue), filamentary edges and cluster nodes (red).}
    \label{fig:spine_filas_walls_surfaces_zoom}
\end{figure*} 

\subsection{Cosmic Spine: Filaments and Walls}
The final step in the SpineWeb method is to identify and label the voxels 
that correspond to the spine (filaments and clusters) and walls, following the {\it SpineWeb} criteria specified in 
equation~\ref{eq:neighbour_morphology}. 

An insightful impression of the intricacy of the full three-dimensional network of filaments 
and walls is presented in fig.~\ref{fig:spine_filas_walls_surfaces}. The top two frames show 
the wall-like (blue) and filamentary (red) regions separately. Clearly outstanding is 
the percolating nature of the filamentary network and the complex of connecting sheets. 

\begin{figure*}[!ht]
  \centering
  \includegraphics[width=0.49\textwidth,angle=0.0]{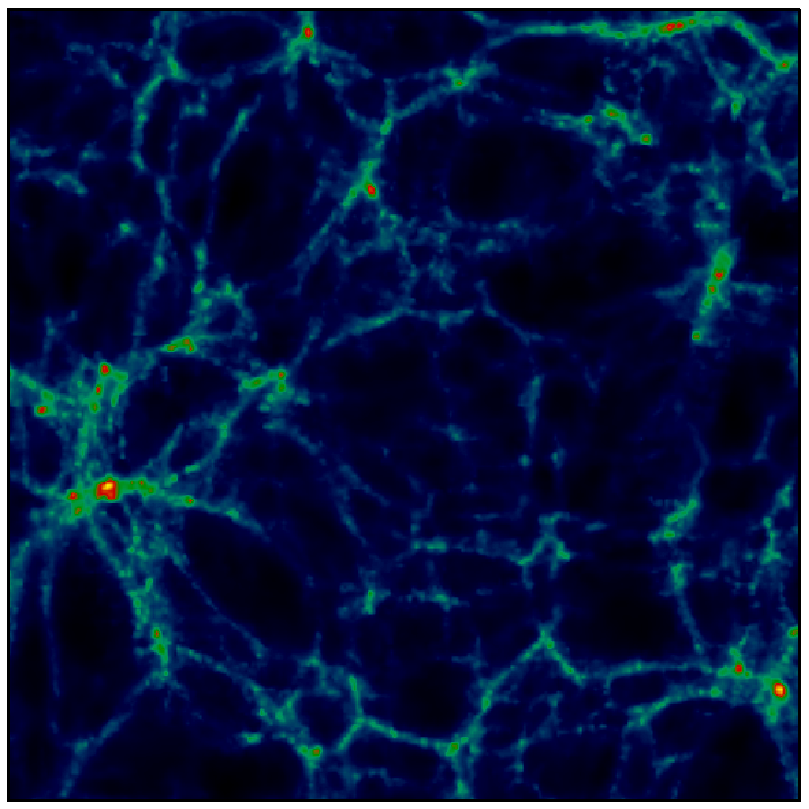}
  \includegraphics[width=0.49\textwidth,angle=0.0]{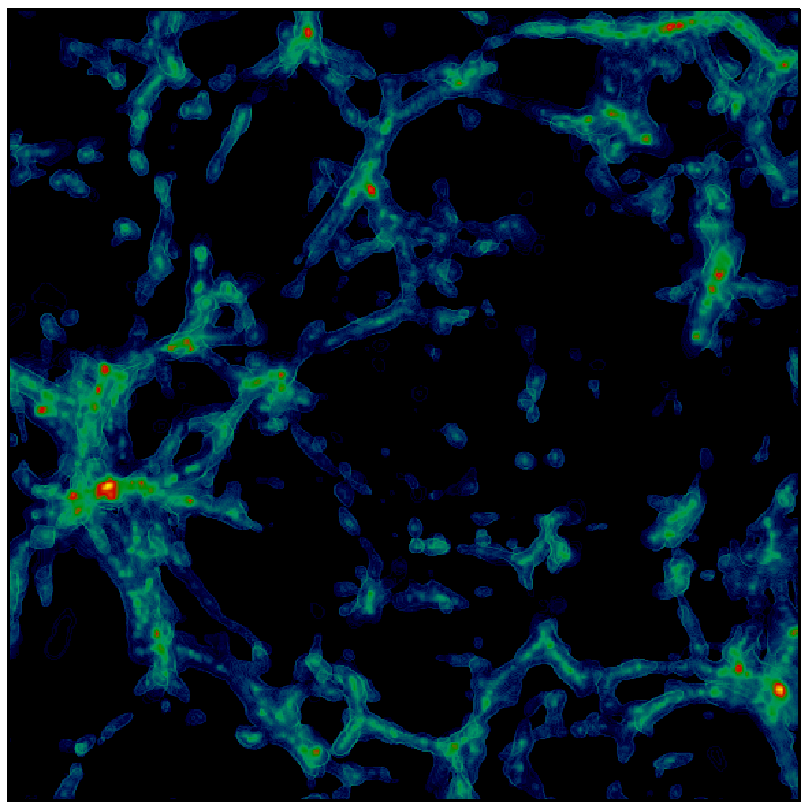}
  \includegraphics[width=0.49\textwidth,angle=0.0]{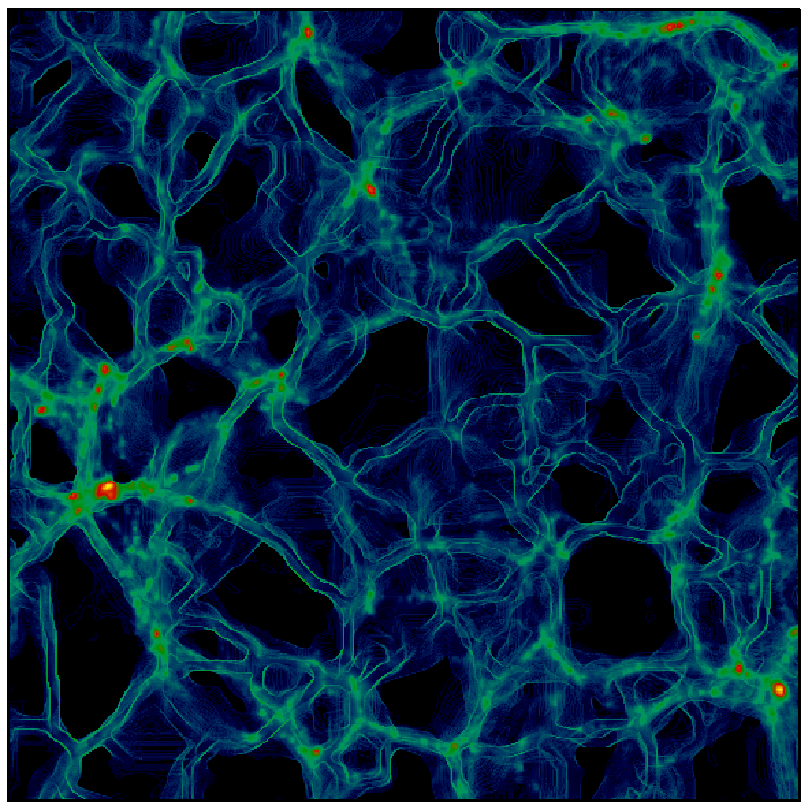}
  \includegraphics[width=0.49\textwidth,angle=0.0]{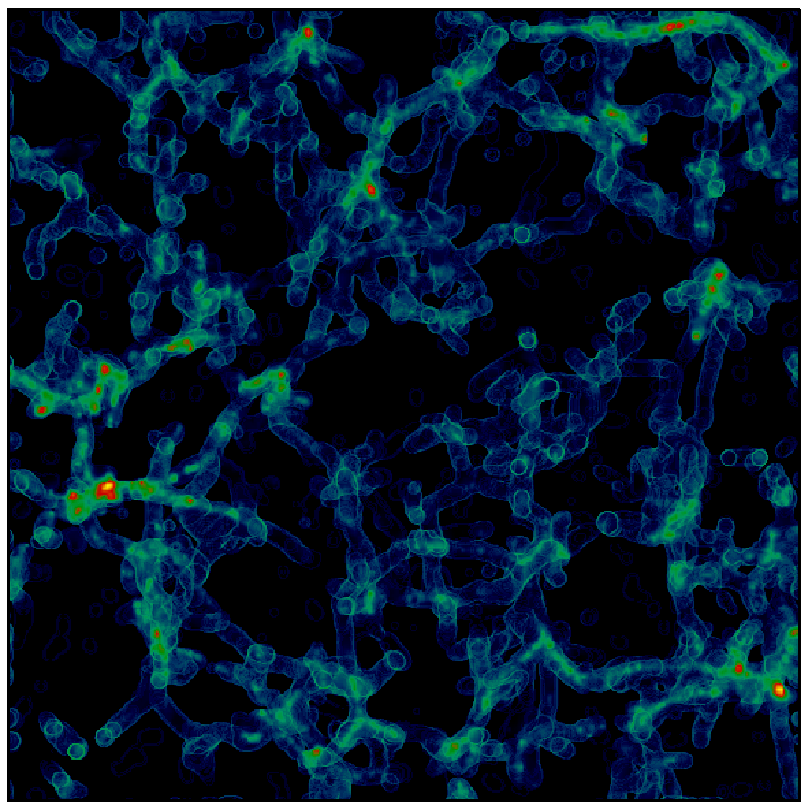}
    \caption{Filaments and walls identified with the SpineWeb algorithm. Top left: Volume rendering of the density field inside a 
             subbox of the simulation. Top right: density field contained inside an isosurface at $\delta = 1$. Only the 
             density inside the isocontour is plotted. Bottom left: density field and semi-transparent isosurfaces delineating 
             walls. Bottom right: density field and semi-transparent isosurfaces delineating filaments. Both wall and 
             filament mask have been smoothed for visualization purposes with a Gaussian
	     filter of $\sigma=2$ pixels.}
  \label{fig:spine_densi_filament_walls_ghost}
\end{figure*}

The appearance of a uniform width, at places considerably in excess of the local width of the 
density field contours, is a result of our choice to show, for visualization purposes, a uniform 
smoothed outline. The plotted isosurface of both walls and filaments is obtained by filtering the 
mask defined by all wall voxels with a Gaussian kernel of $\sigma=2$ voxels. The Gaussian smoothing 
radius corresponds roughly to the average width of $\approx 2h^{-1}$Mpc of filaments and walls, as 
we found in a previous study \citep{Aragon07}. One may get an impression of the corresponding 
variation in density and width along the spinal structures by inspecting the two bottom frames of 
fig.~\ref{fig:spine_densi_filament_walls_ghost}, where we have superimposed density contour 
levels onto the embedding spinal contours (left: walls; right: filaments). 

The top-left panels in figure ~\ref{fig:spine_filas_walls_surfaces} shows a network of complex sheets. 
Instead of a regular ``planar'' geometry, on small scales the walls have a curved appearance marked by an irregular surface. To a 
considerable extent this reflects their inhomogeneous internal mass distribution, itself a result 
of their hierarchical buildup. The irregular convoluted shapes are found 
on all scales, although the walls do have a slightly more regular semi-planar geometry on larger scales. 

Also the filamentary structures reflect their inhomogeneous internal mass distribution, even though 
the applied smoothing has tended to diminish the contrast between e.g. massive clumps and tenuous moderate 
or lower density parts of the filaments. As a result, over most of its outline the filamentary edges look 
semi-linear. Nonetheless, occasionally we can recognize rather twisted configurations, and at numerous 
locations we can recognize the bulging presence of massive clusters. Taking the nodal junctions as the 
endpoints demarcating an individual filament, a first analysis shows that short filaments tend to be more 
straight than longer ones. This is entirely in line with the trend predicted by the Cosmic Web theory, 
and is in agreement with a corresponding analysis of N-body simulations \citep[][]{Bond96,Colberg05, Nicholas10}.

\subsection{Morphological Connections}
The close mutual relationship between the walls and filaments is immediately clear when inspecting the superposition 
of the wall-like and filamentary web in the bottom (right) frame of figure~\ref{fig:spine_filas_walls_surfaces}. With 
the filaments defining an interconnected web, the walls fill the spaces in between the filaments, together forming 
a ``watertight'' complex of membranes surrounding a system of voidlike cavities. The zoom-in onto one specific 
region centered around a particularly intricate branching of filaments emanating from a node, in 
fig.~\ref{fig:spine_filas_walls_surfaces_zoom}, highlights the complex connections that may occur 
in the Cosmic Web. At least four filaments appear to originate from a core region at the confluence 
of five walls. In the branches we can clearly recognize the bulging imprint of massive clusters. The 
zoom-in also nicely shows the small-scale heterogeneity of the walls' surfaces.

\subsection{Cosmic Spine versus Density Selection}
To compare the morphological SpineWeb segmentation with the corresponding density field, 
in figure~\ref{fig:spine_densi_filament_walls_ghost} we have depicted the matter distribution in 
a central slice through the simulation box. The two top panels are images of the density field 
(see discussion sect.~\ref{sec:densmorph}), while the bottom panels present the density field 
within morphologically segregated regions. The two frames at the bottom show the density field 
inside semi-transparent surfaces enclosing the wall features (bottom left) and the filamentary 
features (bottom right).  

\subsubsection{Stereological Considerations}
When assessing the bottom frame of fig.~\ref{fig:spine_densi_filament_walls_ghost}, we have to take into account that 
isolated features observed in these slices are the result of the finite thickness of the slice. Particularly noteworthy 
for the filaments in the righthand frame, 
they are artefacts of the finite width of the depicted slice. Because the orientation of filaments in the cosmic spine with 
respect to the slice is random, their intersection with the slice will differ. Dependent on the intersection angle, it may 
vary from their full length - in case they are nearly entirely embedded within the slice -  to a mere point in case they 
run perpendicular to the slice. The resulting impression is one of a semi-irregular distribution of shorter and longer ``stubs'', 
which indeed we find back in the bottom righthand frame. 

The two-dimensional geometry of walls tends results in linear intersection with the depicted slice. This produces the fully 
percolating network of (intersection) edges seen in the bottom lefthand frame. Note that occasionally the orientation of a 
wall is so favorable that its intersection is not a one-dimensional edge but instead consists of a slab comprising 
a major fraction of the wall. In the most extreme circumstance, the wall is entirely embedded within 
the slice so that it remains visible in its entirety. 

\subsubsection{Morphological Structure of Density Features}
While the first superficial impression might be that the isocontour map of the density field 
(see fig. \ref{fig:spine_densi_filament_walls_ghost} top-right panel)
is richer in detail and structure than the filamentary and sheetlike morphologies in the bottom 
frames, a few important observations need to be made. 

An important contrast between the isodensity contour maps and the filamentary and sheetlike 
networks defined by the SpineWeb procedure is that between the rather discontinuous nature 
of the (thresholded) density maps and the fully percolating spinal structure. This is most 
clearly visible when zooming in regions with large density contrasts, of which filaments 
close to the infall region are a good example. The massive cluster complex visible at the 
left of the box forms a nice illustration. The filamentary extensions connecting to the 
cluster are identified by the SpineWeb procedure (lefthand panel fig.~\ref{fig:spine_densi_filament_walls_ghost}), 
along with the sheetlike membranes of which they form the boundary (righthand panel 
fig.~\ref{fig:spine_densi_filament_walls_ghost}). It would be very challenging for 
traditional density-based filament detection techniques to trace filaments near cluster-filament 
interfaces. The density in the infall regions of clusters tends to increase dramatically, 
rendering a density-based criterion to determine the local morphology rather 
cumbersome\citep{Aragon07}.

\section{Analysis Wall \& Filament Sample}
\label{sec:analysis}
In this section we present a few quantitative measures of the voids, walls and filaments 
extracted by the SpineWeb technique. The results concern the single-scale analysis 
of our $64^3$ particle $\Lambda$CDM simulation described in the previous section 
(see sec.~\ref{sec:lcdm}).

\subsection{Density distribution}
\begin{figure}[!fbt]
  \centering
  \includegraphics[width=0.49\textwidth,angle=0.0]{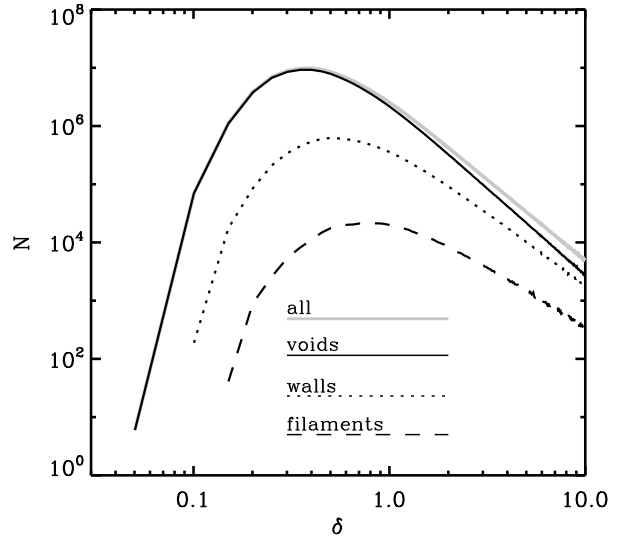}
    \caption{Normalized density distribution for all voxels in the simulation box 
    (thick gray line), void voxels (solid line), wall voxels (dotted line) and
	 filament voxels (dashed line).}
  \label{fig:Cosmic_Web_density}
\end{figure}

The density field of the $64^3$ simulation was computed on a regular 512$^3$ grid, 
no smoothing was applied to the filament and wall masks. 
Figure \ref{fig:Cosmic_Web_density} shows the density field distribution computed for the
complete simulation box, voids, filaments and walls. The distribution of densities can be 
roughly described as log-normal with the main difference between 
morphological environments being the position of the peak of the distribution. 
Numerous studies have shown that a gravitationally evolving matter distribution, starting from Gaussian 
initial conditions, tends to attain a lognormal density distribution towards more advanced quasi-linear 
stages \citep[][]{ColesJones91,Neyrinck09,Platen09}. Our results indicate that this remains true for each of 
the individual morphologies.

Voids have the lowest densities followed by walls and
filaments respectively. It is important to note that the network of filaments found by our method
contains also the clusters which act as the nodes of the wall-filament network. This affects the
right tail of the  distribution and the computed moments.  The median is a much better estimator 
than the mean in all cases given the effect of the clustering of matter into halos in our
sampling schema. 

Table \ref{tab:cosmic_web_stats} shows basic statistics of the density field characterized
by the Spine. We present the statistic computed at two different grid resolutions ($256^3$ and $512^3$ voxels) 
as a simple convergence test. 
The numbers in table~\ref{tab:cosmic_web_stats} show that the mean and median densities of the 
different morphologies are largely similar for the two different resolutions. This is quite 
different for the volume occupancy, in particular for the walls, filaments and clusters. This 
is a direct reflection of the resolution-dependent finite thickness assigned to filaments and 
walls, i.e. the voxel size. The volume fraction of voids is less sensitive to the grid resolution. 
We may understand this in terms of the SpineWeb invariants: void occupancy scales by volume, wall 
occupancy by surface area and filaments by length. The latter is therefore most sensitive to 
density field resolution, voids least. 

\begin{table}[!h]
  \caption{voids, filaments and walls: densities and volumes}
  \label{tab:cosmic_web_stats}
  \begin{center}
    \leavevmode
    \begin{tabular}{llll} \hline \hline              
{\bf Structure}        \\
&  $(1+\delta)_{256}$ & & ${\cal F}_{m,256}$ \\
&{\bf mean}& {\bf median} & \\
\hline 
voids     &  0.81  &  0.56  &  0.825 \\
walls     &  1.68  &  0.94  &  0.158 \\
spine     &  3.46  &  1.58  &  0.015 \\
\hline
\ \\
       &  $(1+\delta)_{512}$ & & ${\cal F}_{m,512}$ \\
&{\bf mean}& {\bf median} & \\
\hline
voids     &  0.86  &  0.58  &  0.889 \\
walls     &  1.85  &  0.93  &  0.103 \\
spine     &  4.47  &  1.57  &  0.006 \\
\hline
    \end{tabular}
\caption{Basic statistics of the density field in voids, walls and 
filaments, for density grids of $256^3$ (top) and $512^3$ (bottom) 
voxels. Quoted are the mean and median of the density field, $(1+\delta)$, 
in the individual morphologies, and the volume fraction ${\cal F}_{V,m}$ 
of each of them. } 
  \end{center}
\end{table}

An important observation is also the considerable overlap between the pdf's of the different 
morphologies. While distributions peak at clearly different places, there is  
a large overlap between all density distributions.  This degeneracy in the density 
distributions between morphologies explains why a given isodensity contours does 
not manage to isolate one specific morphology, but will invariably include regions 
belonging to other morphologies too.

For our purpose, most significant are the differences between density distributions inside 
filaments and walls. Perhaps most remarkable is the sizeable overlap between densities in the 
void fields and those in filaments. The density distribution inside voids is almost identical 
to the overall density distribution. This is not surprising given the fact that even though voids 
are extremely underdense they occupy most of the space in the Cosmic Web. The difference between 
both distributions occurs at the high density tails where the clusters lie. 

\subsection{Minkowski-Bouligand dimension}
We performed a preliminary scaling analysis of the identified filamentary and wall-like networks in 
the analyzed $\Lambda$CDM N-body simulation. To this end, we have determined for each of these 
networks the Minkowski-Bouligand dimension $D_{\textrm{\tiny{MB}}}$ -- or box counting dimension -- 
formally defined as:
\begin{equation}\label{eq:fractal_dimension}
D_{\textrm{\tiny{MB}}}\,=\,\lim_{\epsilon \rightarrow 0} \frac{\log N(\epsilon)}{\log (1/\epsilon)}.
\end{equation}
\noindent In this expression, we count the number $N(\epsilon)$ of boxes of (infinitesimal) size 
$\epsilon$ required to fill or cover the set of points belonging to the filamentary or wall-like 
web. In practice, we divide the simulation box into subboxes of size $s$ and count the number $N(s)$ 
of subboxes that contain at least one voxel labeled as filament or wall. By repeating this evaluation 
for several box sizes, and determining the scaling index $N(s) \propto s^{-D}$, we obtain an estimate 
of the Minkowski-Bouligand dimension. One may visualize this by plotting the count $N(s)$ versus the 
size $s$ if the boxes, preferentially in a logarithmic diagram. If indeed characterized by a single 
fractal dimension, the resulting curve would be characterized by one slope. In practice, the 
structural patterns tend to be more complex, manifesting itself in scaling curves that cannot 
be characterized by a single uniform slope. 

\begin{figure}[!t]
  \centering
  \includegraphics[width=0.49\textwidth,angle=0.0]{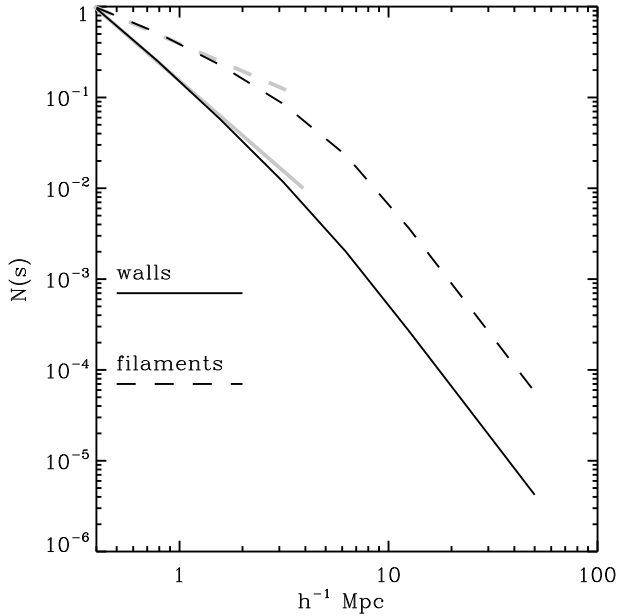}
    \caption{Box-counting dimension of the filamentary spine (black dashed line) and walls (black solid line) of the Cosmic Web.
    For comparison we show the curves for ideal one and two-dimensional objects (gray dashed and gray solid lines respectively).}
  \label{fig:webfracdim}
\end{figure} 
Figure \ref{fig:webfracdim} shows the Minkowski-Bouligand dimension computed for the wall and filament networks.
We also show two (grey) lines with slope -1 and -2 as a reference indicating the cases of a pure one and two-dimensional
objects. The slope of the curves for filaments and walls differ considerably at small scales. 
Filaments behave like one dimensional lines up to scales of $3-4$ $h^{-1}$Mpc after which point the absolute 
magnitude of the slope of the curve increases from -1 to -3 at scales of approximately 10 $h^{-1}$Mpc. 
In the case of the wall network we see a similar behavior with walls having a clear two-dimensional nature at 
scales smaller than $3-4$ $h^{-1}$Mpc. The transition point in the curve of figure \ref{fig:webfracdim} provides a 
good indication of the scale at which filaments and walls start joining each other forming an interconnected network. 
At this point their dimension is no longer 1 (filaments) or 2 (walls) but a higher value reflecting the complexity of the 
network of filaments and walls that form the Cosmic Web. 

\begin{figure*}[!t]
  \centering
  \includegraphics[width=0.49\textwidth,angle=0.0]{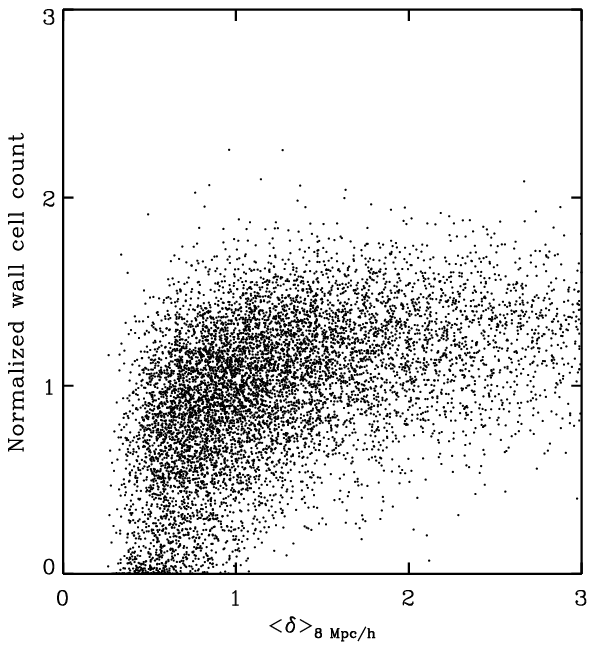}
  \includegraphics[width=0.49\textwidth,angle=0.0]{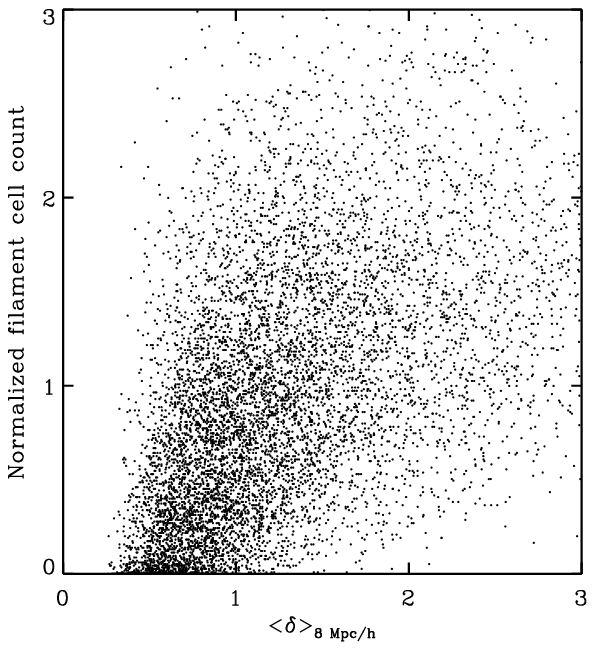}
    \caption{Number of cells labeled as wall (left panel) or filament (right panel) inside boxes of $8$ $h^{-1}$Mpc of side as
    	a function of the mean overdensity inside the $8$ $h^{-1}$Mpc box. The number of cells is normalized with the mean 
        count of all the $8$ $h^{-1}$Mpc boxes. }
  \label{fig:density_cells}
\end{figure*} 

\subsection{LSS complexity and local density}
Another measure of the local complexity of the network of filaments and walls is presented in
figure \ref{fig:density_cells} where we show the mean number of cells labeled as filament or wall inside
boxes of $8$ $h^{-1}$Mpc size as a function of the mean density inside the boxes.  

Low values indicate very simple local configurations while large values reflect complex environments. 
At first glance this may seem straightforward as increasing excursions sets of the density field have a 
similar behavior. However, the filaments and walls we identify are one-voxel thick so their 
voxel count correlates with their length and surface area respectively. For a given fixed volume 
larger counts indicate more intricate filament and wall systems.

We find a trend between the density and the complexity of the environment. Highly dense boxes tend to 
contain more structures than underdense boxes. The regions in the vicinity of massive clusters are a good example 
of complex neighborhoods defined in a locally overdense regions while the large voids define relatively simple 
wall and filament structures. The spread about the clearly visible mean trend is quite substantial. One of the 
main reasons is the restriction in the scale of the analysis, which leads to a confusion of intrinsic scales 
and counting of fainter structures together with more significant ones. A proper multiscale analysis, the 
subject of our following paper, will take this into account. 

\section{Conclusion and future work}
\label{sec:conclusions}
The Spine of the Cosmic Web is the cosmic web's framework, consisting of the network of filaments and walls
and their connections at the clusters nodes. In this study we present a topological technique 
based on the discrete watershed transform of the cosmic density field for the identification and 
characterization of voids, walls and filaments. Our method is closely related to a variety of 
concepts from computational topology, and has a strong mathematical foundation in Morse theory 
of singularities. 

The SpineWeb method is ideally suited for morphological and dynamical studies of the Large Scale Structure. 
Amongst others, it will allow a better insight into the formation and dynamics of the anisotropic filamentary and 
wall-like structures in the Large Scale Universe. Another immediate application is in 
addressing the question whether and which influences the large scale environment has on the halos and galaxies 
that are forming within their realm. 

As a first test of its viability, we applied our method on a set of heuristic Voronoi clustering 
models. The SpineWeb procedure succeeds in reconstructing the original properties of the 
cellular galaxy distribution. In the implementation presented in this work, we effectively restrict 
ourselves to a single spatial scale determined by the voxel scale of the regular grid on which the 
density field is sampled. In a forthcoming paper we will discuss the effect of the multiscale nature of 
the matter distribution. The scale-space formulation of the SpineWeb method will enable us to 
identify fainter features in the density field and establish their connections with other 
objects into a truly hierarchical weblike pattern. In other words, it provides an effective 
way towards characterizing the hierarchy of structures in the Cosmic Web.  

A crucial aspect of the watershed transform and of our method is the definition of local
neighborhood. In the case of regular grids the immediate neighborhood of 26 pixels is arguably the 
best option. However, for unmeshed data such as galaxy surveys and N-body simulations, other 
neighborhood definitions offer a better choice. Among these, the Voronoi contiguous cell defined by 
the Delaunay tessellation of the point distribution represents a promising option. In the 
third paper of this series we will present the result of a Delaunay implementation of the 
Spine method.\\

\section{Acknowledgments}
We would like to thank Bernard Jones for inspiring discussions and insightful comments.
This research was funded by the Gordon and Betty Moore foundation. 

\vfill\eject

\end{document}